\documentclass{emulateapj} \usepackage{psfig} \usepackage{apjfonts}
\usepackage{graphicx}
\usepackage{color}
\newcommand\beq{\begin{equation}}
\newcommand\eeq{\end{equation}}
\newcommand\lsim{\mathrel{\rlap{\lower4pt\hbox{\hskip1pt$\sim$}}
        \raise1pt\hbox{$<$}}}
\newcommand\gsim{\mathrel{\rlap{\lower4pt\hbox{\hskip1pt$\sim$}}
        \raise1pt\hbox{$>$}}}

\begin{document}

\title{Signatures of photon-axion conversion in the thermal spectra and polarization of neutron stars}

\author{Rosalba Perna\altaffilmark{1},Wynn C.~G. Ho\altaffilmark{2}, 
Licia Verde\altaffilmark{3}, Matthew van Adelsberg\altaffilmark{4}, 
Raul Jimenez\altaffilmark{3}}
\altaffiltext{1}{JILA and Department of Astrophysical and Planetary Science,
University of Colorado at Boulder, 440 UCB, Boulder, CO, 80304, USA}
\altaffiltext{2}{School of Mathematics, University of Southampton, Southampton, SO17 1BJ, UK}
\altaffiltext{3}{ICREA and ICC, University of Barcelona (IEEC-UB), Spain}
\altaffiltext{4}{Center for Relativistic Astrophysics and School of Physics Georgia Institute 
of Technology, Atlanta, Georgia 30332, USA}

\begin{abstract}

Conversion of photons into axions under the presence of a strong magnetic
field can dim the radiation from magnetized astrophysical objects.
Here we perform a detailed calculation aimed
at quantifying the signatures of photon-axion conversion in the
spectra, light curves, and polarization of neutron stars (NSs).
We take into account the energy and angle-dependence of the 
conversion probability and the surface thermal emission from NSs.
The latter is computed from magnetized atmosphere models that
include the effect of photon polarization mode conversion due to
vacuum polarization. The resulting spectral models, inclusive of the
general-relativistic effects of gravitational redshift and light
deflection, allow us to make realistic predictions for the effects
of photon to axion conversion on observed NS spectra,
light curves, and polarization signals.  We identify unique signatures
of the conversion, such as an {\em increase} of the effective area of
a hot spot as it rotates away from the observer line of sight. For a star
emitting from the entire surface, the conversion produces
apparent radii that are either larger or smaller (depending on
axion mass and coupling strength) than the limits set 
by NS equations of state.  For an emission region that is observed
phase-on, photon-axion conversion results in an inversion of the plane
of polarization with respect to the no-conversion case.  While the
quantitative details of the features that we identify depend on
NS properties (magnetic field strength, temperature) and axion
parameters, the spectral and polarization signatures induced by
photon-axion conversion are distinctive enough to make NSs very
interesting and promising probes of axion physics.

\end{abstract}


\section{Introduction}

Axions with low mass (below 1~meV) are a direct prediction of the
solution for the strong CP violation problem. Peccei \& Quinn (1977) proposed axions as
Pseudo-Goldstone bosons arising from the spontaneous breakdown of the
U(1) symmetry.  The Peccei-Quinn axion occupies a narrow region in the
axion parameter space of mass and coupling strength. However, axions
(or pseudo-scalar particles) can exist with more generic values for
their mass and coupling strength, and are one type of
dark matter candidate (e.g. Arvanitaki et
al. 2009). This motivates efforts to constrain axion
properties even if the parameters do not reach into the Peccei-Quinn
regime.

Several experimental avenues have been used to detect axions
and constrain their properties.  These fall into two
categories: ground-based experiments and
cosmological-astrophysical efforts.
Current limits on the axion mass
($10^{-6} < m_a/{\rm eV} < 10^{-3}$) come from
the latter approach. The cooling rate 
observed in supernova 1987A implies a direct axion mass upper limit:
if axions are more massive, then cooling
would be predominantly due to axions rather than neutrinos
(Eidelman et al. 2004). The lower limit comes
from cosmological considerations: if axions constitute dark matter, then they must be cold
and not overclose the Universe (Preskill et al. 1983).
Constraints on the coupling constant can be obtained with both
astrophysical observations and ground-based experiments. The existence
of horizontal branch stars with extended morphology, that is,
stars with a range of surface temperatures,
constrains the coupling constant of axion-photon interaction to be $g_{a} <
10^{-10}$~GeV$^{-1}$ (Raffelt 2008).  Ground-based experiments also place
similar constraints, and future experiments are anticipated that will
obtain stronger limits (Andriamonje et al. 1983; De Panlis et al
1987; Wuensch et al. 1989; Asztalos et al. 2004, 2010).

Recent developments in understanding radiation dimming
(Lai \& Heyl 2006; Jimenez et al. 2011) and modification of the
polarization pattern (Gill \& Heyl 2011) have put
further constraints on the $m_a - g_{a}$ plane. Limits on cosmological
radiation dimming have also been used to place constraints on much
lighter axion masses (Avgoustidis et al. 2010).

The potential use of neutron stars (NSs) as probes of axion
parameters has been noted by a number of authors (e.g., Lai \& Heyl
2006; Chelouche et al. 2009; Jimenez et al. 2011). In particular,
occultation and eclipsing or transiting of a companion and
dimming of the spectrum can potentially produce
observable signatures.  However, the phenomenon of occultation (when a
background object passes through the influence of a NS magnetic field)
was deemed to have too low a probability to be astrophysically
relevant.  In addition, there are no known binary systems involving
a NS that are detached enough to yield a clean constraint (Jimenez
et al. 2011).  On the other hand, Lai \& Heyl (2006) showed how the
thermal X-ray spectrum of a NS (modeled as simple blackbody emission)
can be modified in a significant way by the presence of axions.
Spectral features due to axion-photon oscillations have been
discussed by Chelouche et al. (2009), and shown to be relevant for
highly magnetized NSs in the sub-mm wavelength range for an
observationally interesting range of axion parameters.

In this paper, we focus on the soft X-ray band in which NSs are
routinely observed.  The goal of our work is to perform a detailed
calculation of the observable effects that the presence of axions
would have on NS spectra, light curves, and polarization.  Our
analysis goes beyond previous computations in the literature for the
subject by coupling the calculation of photon-axion conversion
probabilities with detailed models of magnetized atmospheres and
general relativistic models for emission from the NS surface.  For the
local surface brightness of the NS, we use accurate magnetized
atmosphere models, which yield the angular and energy-dependent photon
intensities in the two photon polarization modes.  These models
include the effect of mode conversion due to the vacuum polarization.
The local intensity is then imported into a code for computing spectra
and light curves from finite emission regions on the NS surface. The
analysis incorporates the effects of gravitational redshift and
gravitational light deflection and allows for arbitrary viewing
geometries.  The resulting spectra are then modified by the
photon-axion conversion process, accounting for both the energy and
angular dependencies of the conversion probability.  {Note that we
  solve the conversion between photon modes and the conversion between
  axions and photons independently, which is shown by Lai \& Heyl
  (2006) to be a very good approximation for the regimes we explore in
  this paper.}  With our calculations, we are able to produce
realistic models for the spectra, light curves, and polarization
signal of NSs, and use these to identify specific features that carry
the telltale signs of photon-axion conversion.  We discuss these signs
in the context of NS observations in \S5.2.

Our paper is organized as follows: In \S2, we discuss the formation of the
NS thermal spectrum within a magnetized atmosphere, placing particular
emphasis on the emergence of two photon polarization modes. In \S3, we summarize
the calculation of the photon-axion conversion probability for a magnetized object.
The computation of the `observed' thermal spectrum, light curves, and
polarization with and without photon-axion conversion is described in \S4, and
the results are presented in \S5. We summarize our findings in \S6. 

\section{Photon propagation in the atmosphere of magnetized neutron stars}

In this section, we describe the main properties of magnetized
NS atmospheres and the processes that influence photon
propagation in them.  Axions do not directly couple to the processes
described in this section.  However, the details presented below are
important for calculating the polarized radiation emerging
from the NS atmosphere.  As discussed in \S3, axions couple
only to photons that are linearly polarized in the plane formed by the
directions of propagation and the magnetic field; the physics
of photon propagation in the NS atmosphere
determines the relative fraction of such photons.

Radiation emerging from the hot surfaces of magnetized NSs is the
result of radiative transfer through a layer of ionized plasma.
Due to the strong gravitational field
of the NS, heavy elements settle quickly
(Alcock \& Illarionov~1980; Brown et al.~2002),
and this atmospheric layer is composed of light elements,
unless nuclear burning on the NS surface consumes the light
elements (Chang et al. 2010, and references therein).
Only recently have self-consistent magnetic atmosphere models for
partially ionized
hydrogen and mid-$Z$ elements been constructed (see Mori \& Ho~2007;
Ho et al.~2008, and references therein for details;
see also Suleimanov et al.~2009).
For simplicity, we focus our calculations on fully ionized hydrogen
atmospheres (see below).
In this case, the scale height is 
\beq
H_\rho=\frac{2kT_s}{m_p g \cos\delta}=
\frac{1.65}{ \cos\delta}\left(\frac{T_s}{10^6\, {\rm K}}\right)
\left(\frac{g_*}{10^{14}\,{\rm cm\,s^{-1}}}\right)^{-1}\,{\rm cm}, 
\label{eq:Hrho}
\eeq

\noindent where $T_s$ is the effective temperature of the star,
$g_*$ is the gravitational acceleration,
and $\delta$ is the angle that the ray makes with the surface normal.

In a NS atmosphere, radiation propagates in two distinct polarization
modes: the extraordinary (X) mode and ordinary (O) mode, which are
very nearly linearly
polarized perpendicular and parallel, respectively, to the plane
formed by the directions of propagation and the magnetic field.
The X mode absorption opacity is reduced by a factor of
$(E_{Be}/E)^2$ relative to the O mode opacity, where the electron cyclotron
energy is $E_{Be}=115.8 B_{13}$~keV and $B_{13}=B/10^{13}\mbox{ G}$.
Thus the X mode
radiation decouples from deeper, hotter layers in the NS atmosphere
than the O mode radiation. For rays propagating at intermediate angles
$\theta_{p}$ relative to the magnetic field, the net emission
is significantly polarized and dominated by the X mode.

To produce surface emissivities for the X and O photon modes, we use
the model of van Adelsberg \& Lai (2006), which quantitatively
incorporates the effects of vacuum polarization on the radiative
transfer for fully ionized atmospheres with an external magnetic field
parallel to the axis of symmetry. This model is reasonably
accurate for temperatures $T_s\ga 10^6$~K. At lower temperatures,
partial ionization of atomic species becomes important, and partially
ionized models are needed (see Ho et al. 2008, and references therein).
Therefore, we use the fully ionized model only in the high
temperature regime. It should be noted,
however, that this is the effective temperature of the star, and that
the measured value at the observer is lower due to
gravitational redshift. Colder stars are less interesting for the
purpose of this study: First, being much dimmer,
their measured spectra are generally of lower statistical significance
for detailed spectral studies. Second, and more importantly, 
since the fraction of O-mode photons
increases with temperature, spectra of colder stars are
less sensitive to axion effects (see the discussion below).

For magnetic field strengths $B\sim 4\times 10^{13}$~G, there is a
significant contribution from vacuum polarization to the dielectric
properties of the medium (Adler~1971; Tsai \& Erber~1975).
A ray traversing the density gradient of a magnetized
NS atmosphere will eventually encounter a layer of the
medium in which the plasma and vacuum contributions to the dielectric
tensor are of the same magnitude, leading to a resonance. For the
models studied in this paper, the resonance occurs at lower density
than the X and O mode decoupling depths (where the optical depth
$\tau_\nu\approx 1$) for most photon energies
and propagation angles. In addition, the integrated opacity across the
vacuum resonance is negligible at these field strengths.
A discussion of vacuum resonance effects on the mode opacities is given by
Ho \& Lai~(2003).  As shown by Lai \& Ho (2002) and Lai \& Ho
(2003a), there is coherent mixing of the modes at the resonance,
analogous to the Mikheyev-Smirnov-Wolfenstein (MSW) effect for neutrino
oscillations (see, e.g., Bahcall~1989; Haxton~1995).
Using the geometric optics approximation, and
neglecting damping terms in the dielectric tensor  (which only affect the
width of the resonance),
the amplitudes of the two modes, $A_O$ and $A_X$, evolve according to 
\beq i{d\over dz}\left(\begin{array}{c}
    A_O\\
    A_X \end{array}\right) =\frac{\omega}{2}
\left(\begin{array}{cc}
   2+\sigma_{11}   & \sigma_{12}  \\
    \sigma_{21}  &   2+\sigma_{22}\end{array} \right)
\left(\begin{array}{c} A_O \\ 
                       A_X                       
\end{array}\right),
\label{eq:evolmodes}
\eeq

\noindent where (e.g. Lai \& Heyl 2006)  
\begin{eqnarray}
\label{eq:matrix1}
\sigma_{11}\; = \;[\xi\;-\;v_e]\sin^2\theta_p\;-\;\frac{v_e}{1\;-\;u_e}\cos^2\theta_p\,,\noindent\\
\label{eq:matrix2}
\sigma_{22}\;=\; -\chi\;\sin^2\theta_p\;-\;\frac{v_e}{1\;-\;u_e}\,,\noindent \\
\label{eq:matrix3}
\sigma_{21} \;=\;-\sigma_{21}\;=\; i\frac{v_e\,u_e^{1/2}}{1\;-\;u_e}\cos\theta_p\;.
\end{eqnarray}
In equations (\ref{eq:matrix1})-(\ref{eq:matrix3}), $\theta_p$ is the angle between the magnetic field and
the photon direction (defined to be the $z$ direction),
$\omega$ is the photon frequency,
$v_e=(\omega_{pe}/\omega)^2$ and $u_e= (\omega_{ce}/\omega)^2$,
where $\omega_{pe}$ and $\omega_{ce}$ are the electron plasma and cyclotron frequencies,
respectively. 
The $B$-field dependent functions $\xi(b)$ and $\chi(b)$ are (Potekhin et al. 2004)

\beq
\xi(b) = \frac{4{\alpha_f}\,b^2\, (1\;+\;1.2b)}{45\pi\;[1\;+\;1.33b\;+\;0.56b^2]}\;,
\label{eq:xi}
\eeq

\noindent and
\beq
\chi(b) = \frac{7{\alpha_f}\,b^2}{45\pi\;[1\;+\;0.72b^{5/4}\;+\;(4/15)b^2]}\;,
\label{eq:chi}
\eeq

\noindent where ${\alpha_f}$ is the fine structure constant and $b$ is the
magnetic field strength in units of the critical field $B_c=m_e^2
c^3/(e\hbar)=4.414\times 10^{13}$~G.
The equations above assume that $v_e\ll 1$ and $u_e\gg 1$.
Since $v_e\sim 10^{-3}(\rho/1\mbox{ g cm$^{-3}$})$ and $u_e\sim 10^4(B/10^{13}\mbox{ G})$ at 1~keV,
these constraints are easily satisfied throughout the soft X-ray band in which NS thermal spectra typically peak.

\begin{figure*}[ht]
\includegraphics[scale=0.9, angle=0]{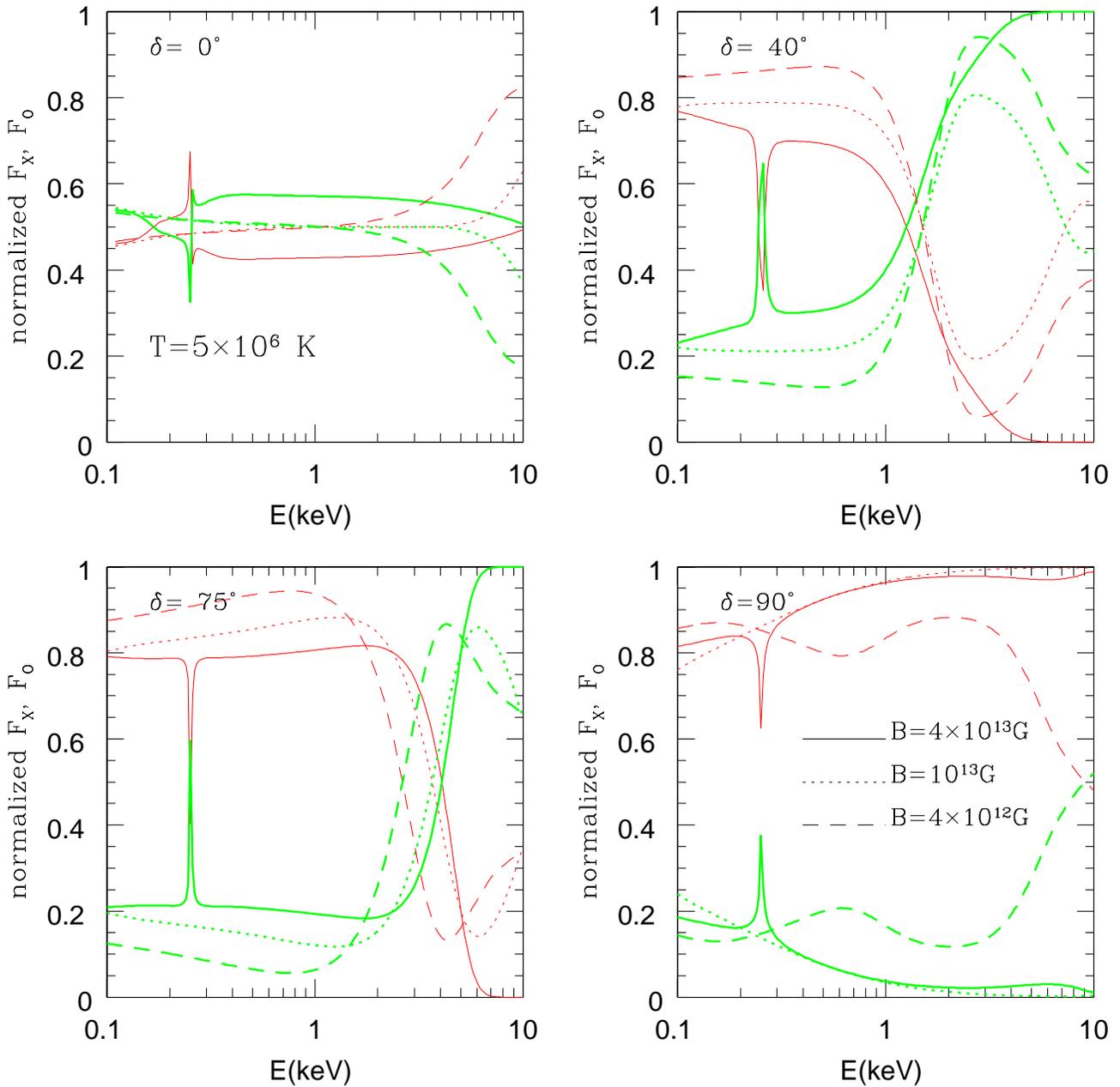}
\caption{Fractional contributions to the flux from the emergent O {\em ( thick/green lines)} and X 
{\em (thin/red lines)} modes
for four angles $\delta$ with respect to the surface normal and three
magnetic field strengths. The effective
temperature is $T_s=5\times 10^6$~K in all cases.
\label{fig:fluxesB}}
\end{figure*}

\begin{figure*}[ht]
\includegraphics[scale=0.9, angle=0]{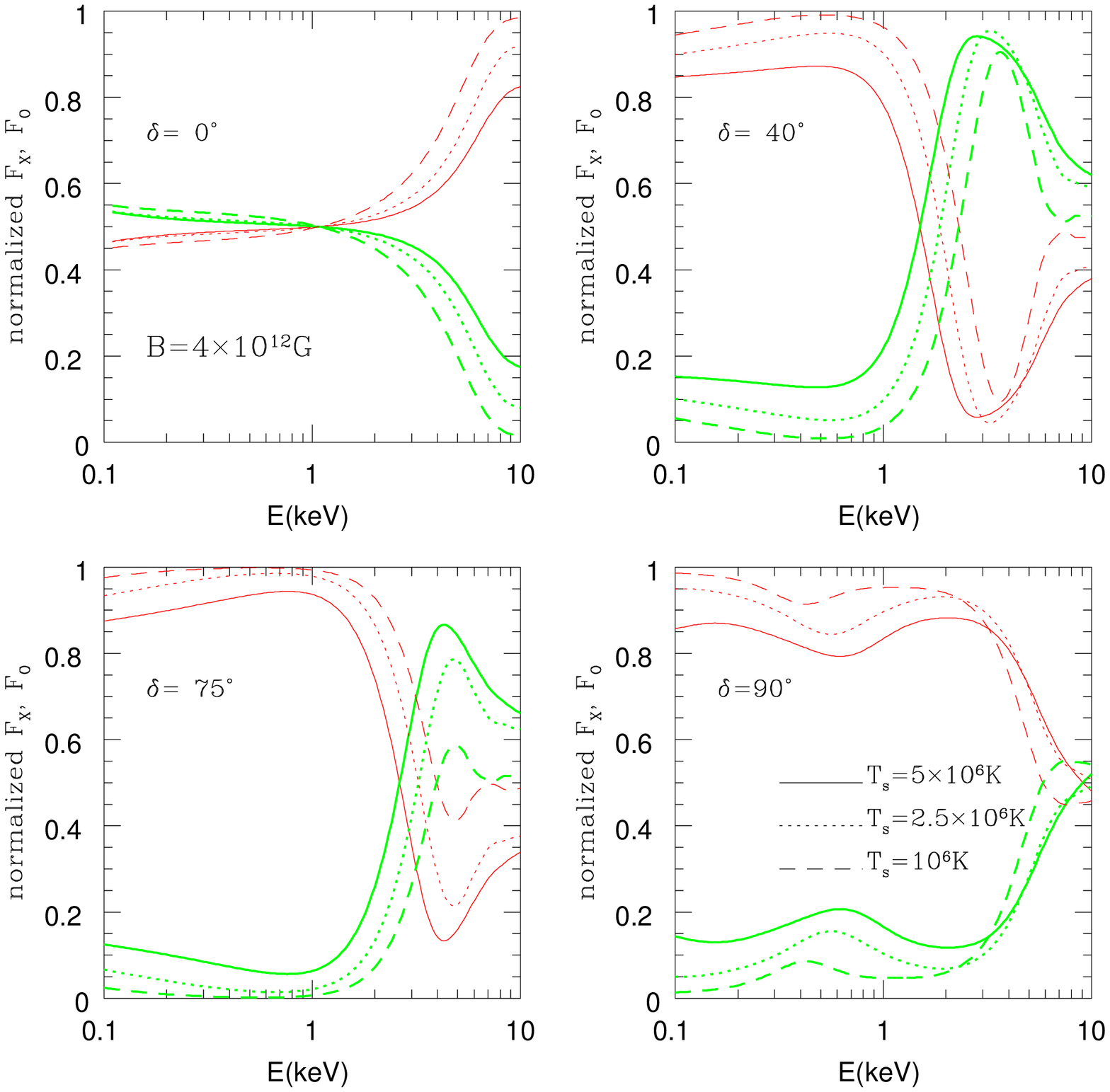}
\caption{Fractional contributions to the flux from the emergent O {\em (thick/green lines)} and X 
{\em (thin/red lines)} modes,
for four angles $\delta$ with respect to the surface normal and three
effective temperatures of the star.
The magnetic field strength is $B=4\times 10^{12}$~G in all cases. 
\label{fig:fluxesT}}
\end{figure*}

Equation~(\ref{eq:evolmodes}) determines the evolution of the photon mode
amplitudes through the atmosphere
(see Ho \& Lai~2003; Lai \& Ho 2003, for details).
The amplitudes are related to the
specific intensities of each mode by $I_{X,O} \propto |A_{X,O}|^2$.
As the photons encounter the resonance, the polarization mode 
ellipticities, $K_{X,O}$ $(\propto E_x/E_y)$,
experience a discontinuity at the resonance. As discussed in
Ho \& Lai (2003), an alternative description of the polarization state uses
ellipticities $K_{\pm}$, which vary continuously
through the vacuum resonance. If the variation of the atmosphere
density profile is gradual enough, an adiabatic condition is
satisfied (see below), and the mode properties are fixed along the curves
described by $K_{\pm}$. Before the resonance, a $K_-$ photon
corresponds to an X mode polarized photon, while a $K_+$ photon
corresponds to an O mode photon.  After the resonance, the
correspondence between $K_{\pm}$ and $K_{X,O}$ switches, so that under
adiabatic conditions the character of each modes changes. This is
analogous to the correspondence between flavor and mass eigenstates in the
MSW mechanism for neutrino mixing.
The adiabatic condition is set by evaluating
the coefficient matrix in Eq.~(\ref{eq:evolmodes}) at the
resonance and setting the magnitude of the diagonal term equal to the
off-diagonal term (Lai \& Ho, 2002; 2003a). In the non-adiabatic
regime, the diagonal terms have a much larger magnitude than the
off-diagonal terms, leading to a set of decoupled equations for
$A_{X,O}$. In the adiabatic regime, the off-diagonal terms dominate
over the diagonal ones, leading to a coupled set of equations for
$A_{X,O}$ and mixing of the polarization states.

The conversion probability $P_c$ is derived from Eq.~(\ref{eq:evolmodes}) by
setting the radiation completely in one mode (e.g., $A_O = 1$, $A_X =
0$) and evolving the equations to $z\rightarrow\infty$. If the
atmosphere density profile is linear in z, then $v_e$ ($\propto\rho$) is
a linear function of z, and the conversion fraction is given by the
well known Landau-Zener formula (see
Lai \& Ho, 2002; 2003a, and the references therein):

\begin{equation}
\label{eq:PC}
P_C = 1 - \exp\left[-\pi\left(E/E_{\mathrm{ad}}\right)^3/2\right].
\label{eq:pjump}
\end{equation}
The adiabatic energy $E_{\mathrm{ad}}$ is defined by
\begin{equation}
\label{eq:Ead}
E_{\mathrm{ad}} \approx
2.52\left[f_B\tan\theta_p\left|1-\left(E_{Bi}/E\right)^2\right|\right]^{2/3}
H_{\rho}^{-1/3},
\end{equation}
\noindent
where $E_{Bi}$ is the ion cyclotron energy, and $f_B^{-2}$ is a slowly varying
function of the magnetic field (see van Adelsberg \& Lai, 2006, and
the references therein).
\noindent Because the vacuum resonance is very narrow in energy (and hence
density, for a given photon energy and propagation angle), the linear
approximation for the density profile is always satisfied for the
atmosphere models considered in van Adelsberg \& Lai (2006). In
addition, the conversion occurs on a distance scale much less than the
scale height of the atmosphere; thus the conversion process can
be regarded as occurring exactly at the resonance density,
and the asymptotic solution to Eq.~(\ref{eq:evolmodes}), in the form
of the conversion probability Eq.~(\ref{eq:PC}), can be used to treat mode
conversion effects. This was shown to be an accurate approximation in
van Adelsberg \& Lai (2006), and we adopt it here in our calculations of mode conversion.

A photon with energy $E$ encounters the vacuum resonance at a density
\begin{equation}
\label{eq:rhoV}
\rho_V \;\approx\; 0.96\;\left(\frac{E}{1\,{\rm keV}} \right)^2\; 
\left(\frac{B}{10^{14}\,{\rm G}}\right)^2\; f_B^{-2}~\mbox{g cm}^{-3}\,.
\end{equation}
For a given energy, $\rho_V$ is the density at which 
the vacuum and plasma contributions are of equal magnitude
(i.e., when $\sigma_{11}=\sigma_{22}$). 
If $E > E_{\rm ad}$, then conversion between the modes at this depth is
effective.  If $I_X$ and $I_O$ are the values of the mode specific
intensities at the resonance, then after conversion they become:
\begin{eqnarray}
\label{eq:Pconv}
I_X'\; = \;\left(1\;-\;P_C\right)\;I_X\; +\; P_C \,I_O\,\nonumber \\
I_O'\; =\; P_C\, I_X\; + \;\left(1\;-\;P_C\right)\;I_O\,.
\end{eqnarray}

This process is automatically incorporated into the models with $B = 10^{13}$~G and
$B = 4\times 10^{13}$~G ($T_s = 5\times 10^6$~K) using the code of van
Adelsberg \& Lai (2006). For models with $B=4\times 10^{12}$~G,
we use the code of Ho \& Lai (2001). This code does not
include partial mode conversion, but we have included the effect
according to the following prescription:
The code of Ho \& Lai~(2001) produces atmosphere profiles
${\rho_d}$ and ${T_d}$ at a set of
discrete points in Thomson optical depth, $\tau_d$, where $d = 1, \dots, D$. Since the
conversion process occurs at lower densities than both the X and O mode
decoupling depths, it does not affect the atmosphere structure and
can be
treated as a modification operating on the emergent specific intensities.
For a
given photon energy and propagation angle, we calculate the vacuum resonance
density using eq.~(\ref{eq:rhoV}), and then locate a grid point $i$, such that
$\rho_i \le \rho_V \le \rho_{i+1}$. We then compute a linear approximation to the
resonance Thomson depth $\tau_V$, using the formula:
\begin{equation}
\tau_V \approx \tau_i + \Delta\tau\left(\rho_V - \rho_i\right)/\Delta\rho
\end{equation}
where $\Delta\tau = \tau_{i+1} - \tau$, and $\Delta\rho = \rho_{i+1} - \rho_i$.
We relate the resonance temperature $T_V$ 
to the Thomson depth and density in a hydrostatic
atmosphere using
\begin{equation}
T_V \approx 1.52\left(\frac{g_*}{10^{14}\;{\rm cm^2}{\rm s}^{-2}}\right)  
\frac{\tau_V}{\rho_V},
\end{equation}
and calculate the atmosphere scale height at the vacuum resonance according
to Eq.~(\ref{eq:Hrho}).
Finally,  the adiabatic energy and mode conversion probabilities are calculated
according to Eqs.~(\ref{eq:PC}) and (\ref{eq:Ead}). The emergent intensities
are then mixed using Eq.~(\ref{eq:Pconv}).

When the surface magnetic field exceeds the quantum critical field
$B_c\sim 4 \times 10^{13}$~G, there are significant vacuum
contributions to the dielectric properties of the medium.  At the
resonance, if the radiation energy is much greater than the adiabatic
energy [Eq.~(\ref{eq:Ead})], there is a high probability of conversion
between the X and O modes.
For magnetic field strengths $B\ga 7\times 10^{13}$~G, the vacuum
resonance lies between the photospheres 
of the X and O
modes (where the mode opacities $\tau_\nu\approx 1$). The modes mix at the resonance with a fraction of X-mode
photons converting into O-mode photons and vice-versa according to the
probability in Eq.~(\ref{eq:pjump}).  The O-mode photons, after
converting into X-mode photons at the resonance, decouple from the
atmosphere.  The X-mode photons that convert to the O-mode,
subsequently interact strongly with the atmosphere, as they now
experience the large opacity for O mode polarization.  Any X-mode
photons that do not convert encounter a large integrated opacity at the 
resonance (under typical conditions; see
Ho \& Lai, 2003).  Thus,
one of the net effects of vacuum polarization is to shift the location of the X-mode
photosphere to the resonance density, which is at lower
density and temperature than the original X-mode photosphere.  As a
result, the high energy spectrum is softer, and the relative fraction of X to O mode photons decreases.  At
magnetic field strengths $B\ll 7\times 10^{13}$~G, radiation in both
modes decouples from the NS atmosphere before encountering the vacuum
resonance.  In this case, mode conversion essentially
switches the photospheres of the two modes, with the O-mode photons
now being produced in the deeper (hotter, denser) layers of
the star. Therefore, if the emission is mostly polarized in the
X mode, for $E>E_{\rm ad}$, the plane of polarization switches to the
O mode after resonant conversion (Lai \& Ho 2003a, 2003b).
The above considerations are particularly relevant for the subsequent
mixing between O-mode photons and axions, as emission which
would otherwise have a small fraction of O mode radiation now
predominantly consists of O mode photons, depending on the magnetic
field strength and NS geometry (see van Adelsberg \& Perna 2009,
for more discussion).

Figure~\ref{fig:fluxesB} shows the fractional contribution to the emergent flux of the X 
and O modes for three values of the field strength, $B = 4\times 10^{13}$~G,
$10^{13}$~G, and $4\times 10^{12}$~G. The effective temperature 
is $T_s = 5\times 10^6$~K in all cases. Each panel shows the results for a separate emission
angle $\delta$ relative to the magnetic field. In general, the X mode
dominates the emergent flux at low energies, except around the ion cyclotron
feature $E_{Bi}$.
According to the formula for $E_{\rm ad}$ [eq.~(\ref{eq:Ead})], there is an angle-dependent
point at which mode conversion becomes efficient and the radiation is composed mostly of
O mode photons. Since $E_{\rm ad} \propto \tan^{2/3}\theta_p = \tan^{2/3}\delta$ for a magnetic field 
perpendicular to the surface, at
larger values of $\delta$, mode conversion is efficient only at high
energies.
Note that as $\theta_p\rightarrow 0$, the properties of the two modes become
similar, and the difference in the opacities decreases.

Figure~\ref{fig:fluxesT} shows the fractional contributions to
the emergent flux of the X mode and O modes, for a fixed
magnetic field strength, $B = 4\times 10^{12}$~G, at three
effective temperatures
$T_s=5\times 10^6$~K, $2.5\times 10^6$~K, and $10^6$~K.
The general trend is
an increase in the O/X flux ratio as the temperature increases. 
This is straightforward to understand:
since a higher temperature atmosphere produces more photons with higher $E$,
the average ratio $E_{Be}/E$ decreases.  Thus, the X-mode opacity becomes more similar
to the O-mode, leading to an increase in the O/X ratio.

\section{Photon-axion conversion in the magnetic field of neutron stars}
\label{section:prob}

In the following we discuss the effect of photon-axion mixing as a
photon beam propagates through the dipolar field of a magnetized NS.
Firstly, we note that an important assumption of our analysis is that
the conversions between the O and X modes and between the O mode and
axion can be treated separately (each as a two-state mixing) rather
than as a coupled three-state mixing.  This assumption is valid as
long as the vacuum resonance and the photon-axion resonance are well
separated. This was shown to be the case by Lai \& Heyl (2006) for
highly magnetized NSs at soft X-ray energies and axion masses $m_a\la
10^{-3}$~eV.  Thus, the assumption of decoupled mixing is satisfactory
for our regime of interest, and it will be adopted in our analysis.

Our formalism for photon-axion conversion in a magnetized medium
follows that of Raffelt \& Stodolski
(1988).  To be consistent with the notation
adopted in the previous section for the two photon polarization states,
we denote the amplitudes of 
photons polarized parallel and perpendicular to the magnetic field as $A_O$ and $A_X$,
respectively. Under the assumption that the length scale over which
the magnetic field varies is much larger than the photon and axion
wavelengths, the evolution of the photon and axion fields
in the $z$ direction, with frequency $\omega$, is given by  

\beq i{d\over dz}\left(\begin{array}{c}
    A_X\\
    A_O\\
    a\end{array}\right) =
\left(\begin{array}{ccc}
    \Delta_\perp & 0 & 0 \\
          0 &  \omega+\Delta_\parallel & \Delta_M \\
     0 &   \Delta_M & \omega+\Delta_a\end{array} \right)
\left(\begin{array}{c} A_X \\ 
                       A_O \\
                       a                       
\end{array}\right),
\label{eq:evolution}
\eeq

\noindent in units with $\hbar=c=1$, where $a$ is the axion field.
The diagonal matrix elements are given
by 
\beq
\Delta_\perp = \,\frac{\omega\,\xi(b)\,\sin\theta_p^2}{2},
\eeq
\beq\;\; \Delta_\parallel = 
\frac{\xi(b)\, \omega \, \sin^2\theta_p}{2},
\;\; \Delta_a = -\frac{m_a^2}{2\omega}\;,
\label{eq:deltapar}
\eeq 
where $m_a$ is the axion mass and $\xi(b)$ is defined by Eq.~(\ref{eq:xi}).
The off-diagonal component is given by 
\beq \Delta_M \; = \;\frac{1}{2}\;g_a\; B(r)\; \sin\theta_p\,
\label{eq:deltaperp}
\eeq
where $g_a$ describes the coupling strength between the O-mode photons
and the axions, and $r$ is the distance from the stellar surface. Since
mixing occurs only between the axion field and the O-mode photon field,
we neglect the evolution of the X-mode field in the discussion that follows.

\begin{figure*}[ht]
\includegraphics[scale=0.5, angle=0]{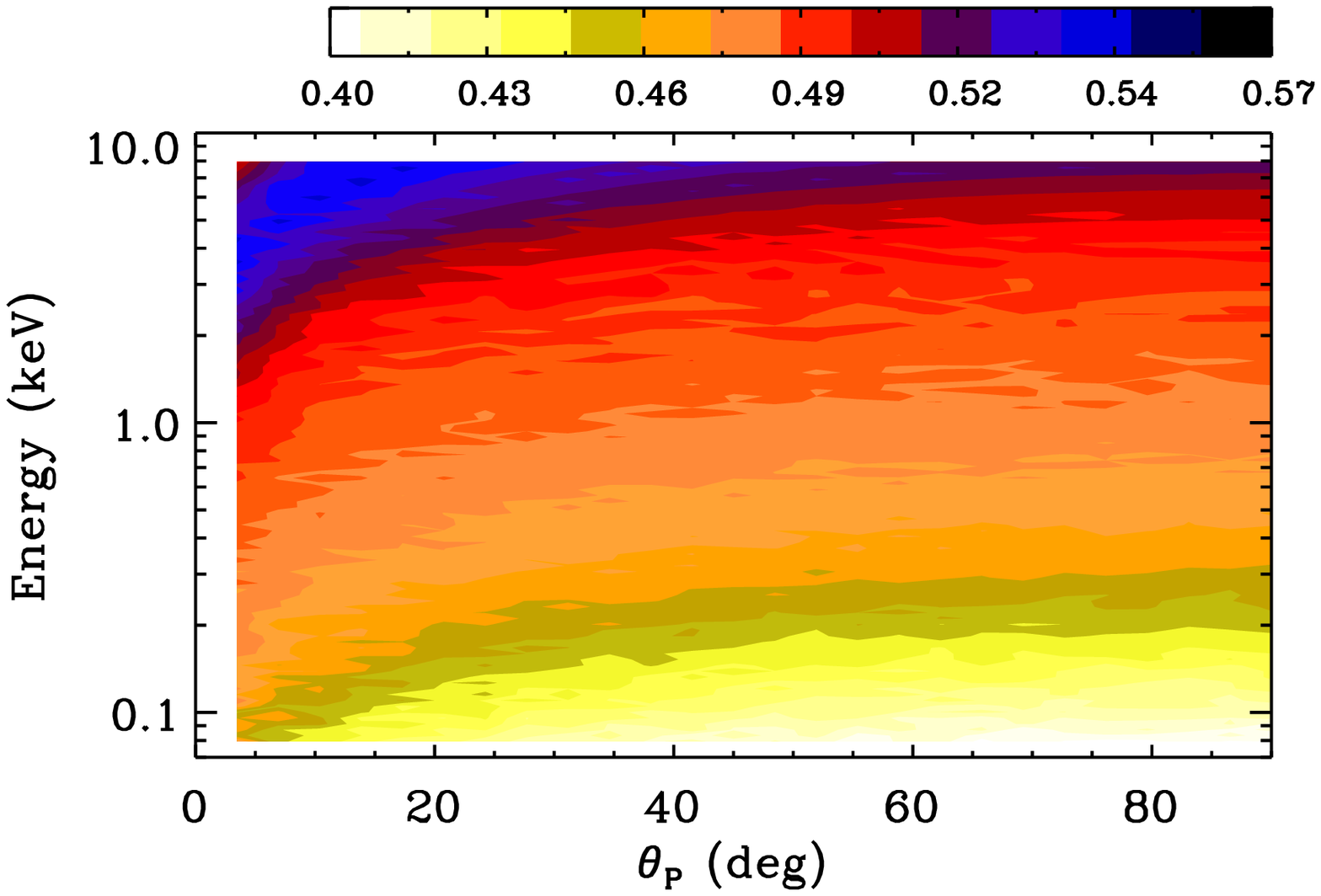} 
\includegraphics[scale=0.5, angle=0]{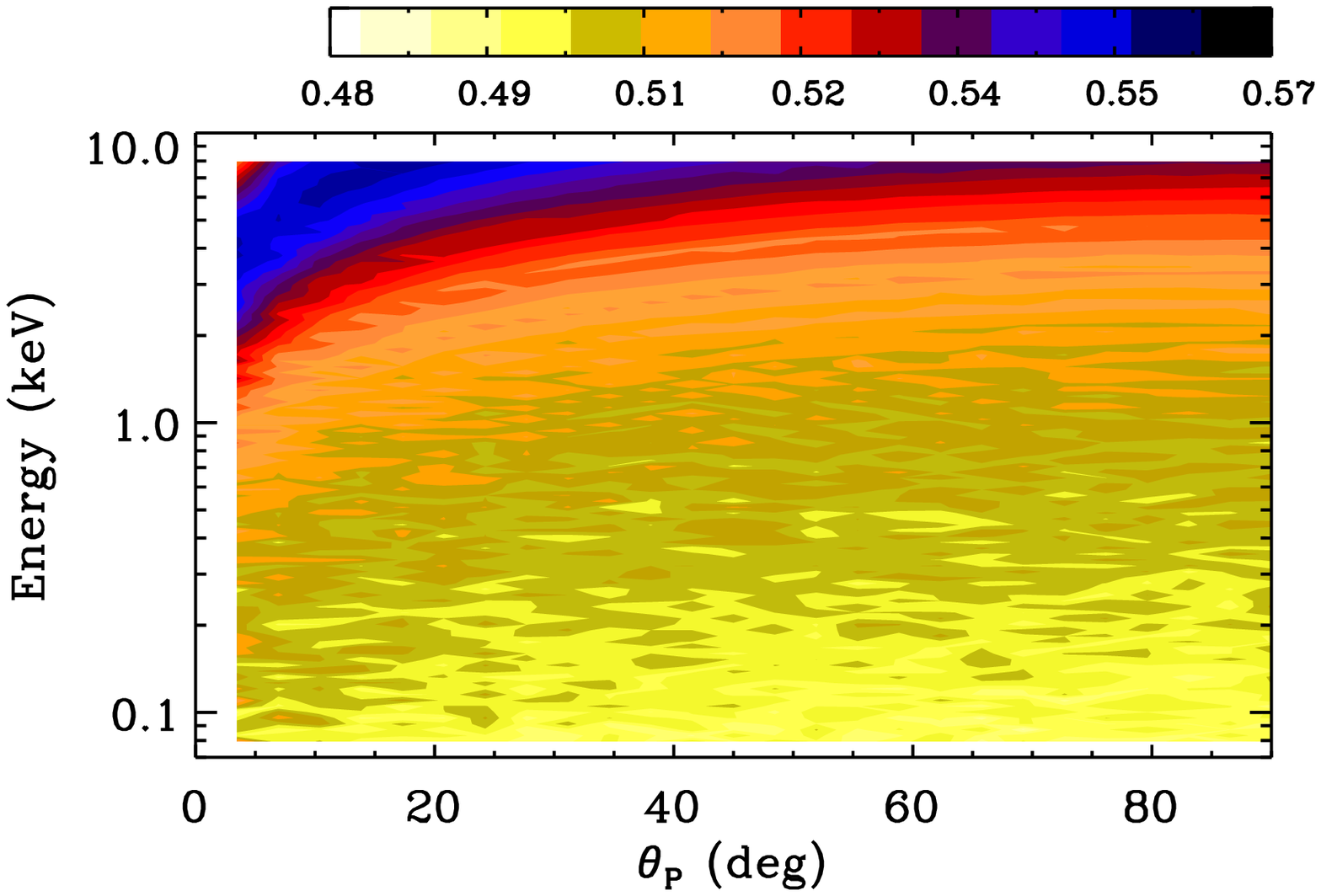} \\
\includegraphics[scale=0.5, angle=0]{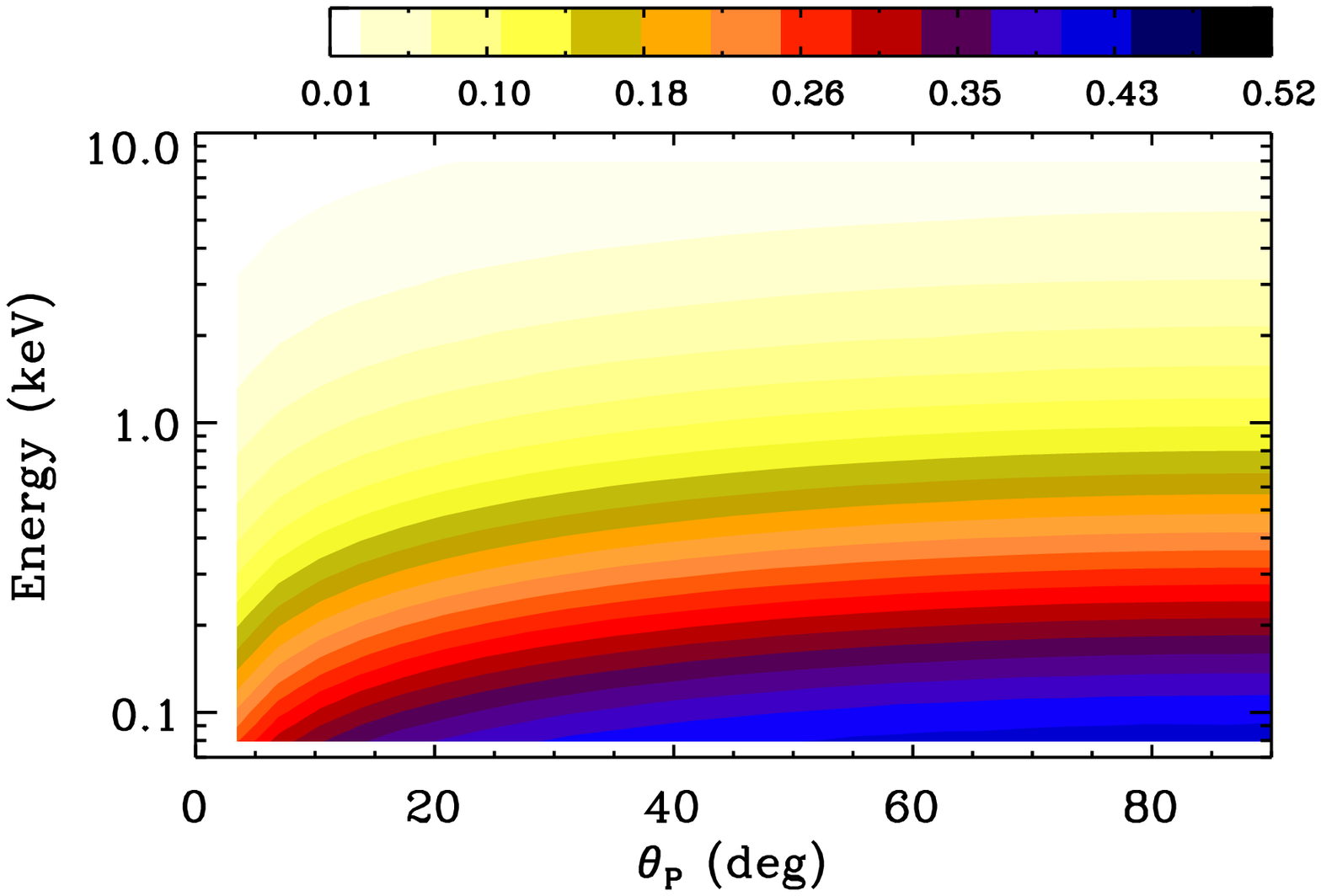} 
\includegraphics[scale=0.5, angle=0]{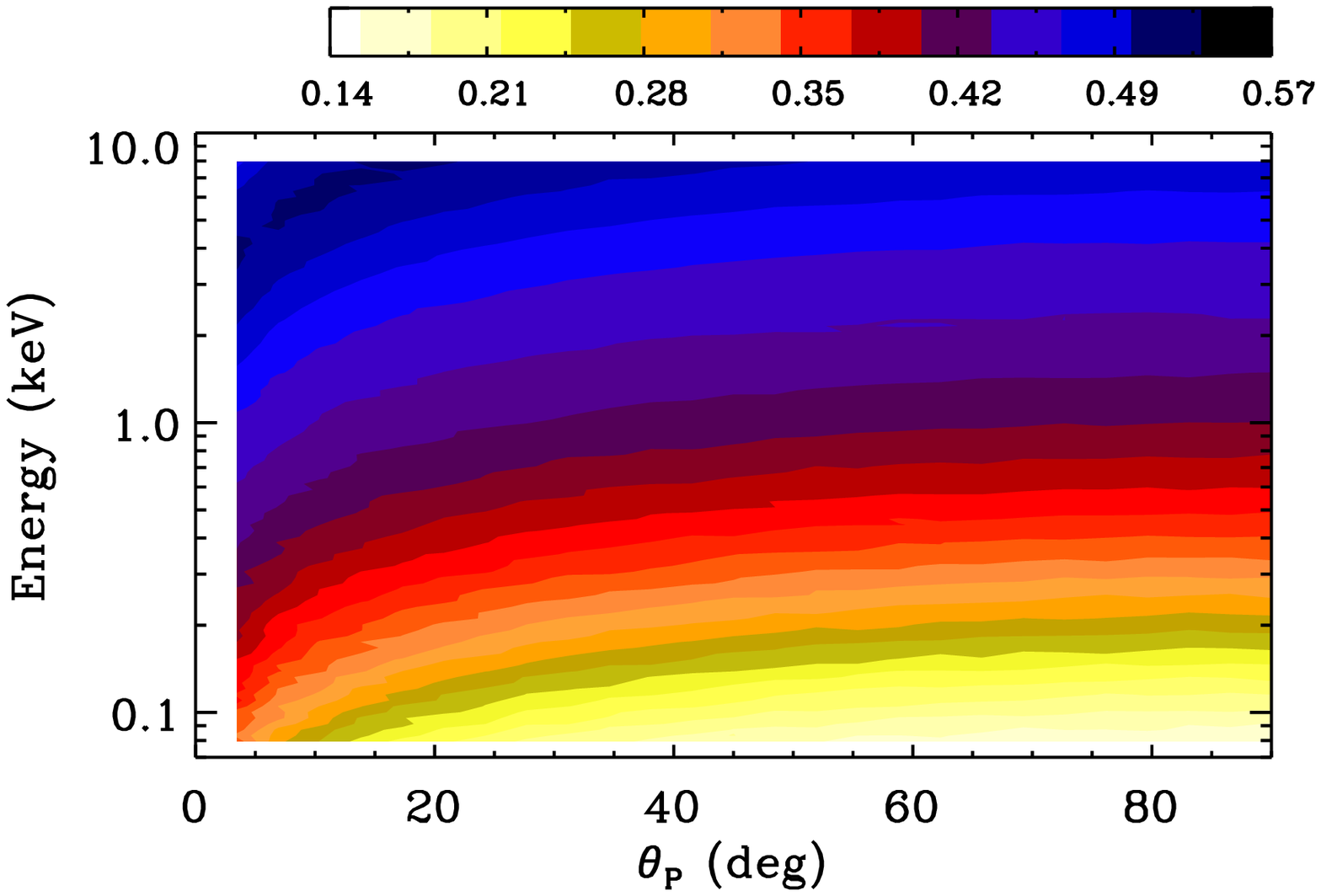} \\
\includegraphics[scale=0.5, angle=0]{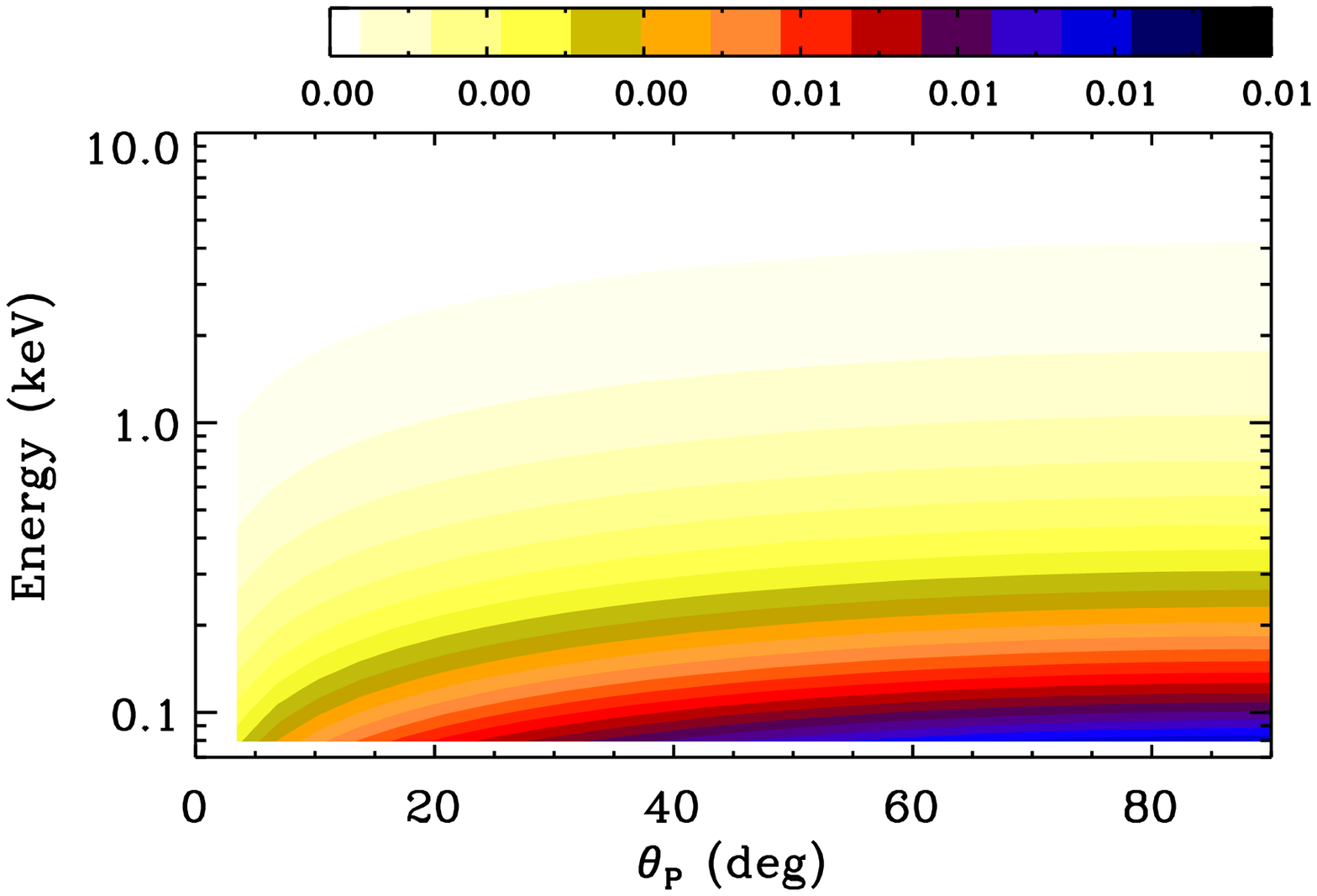} 
\includegraphics[scale=0.5, angle=0]{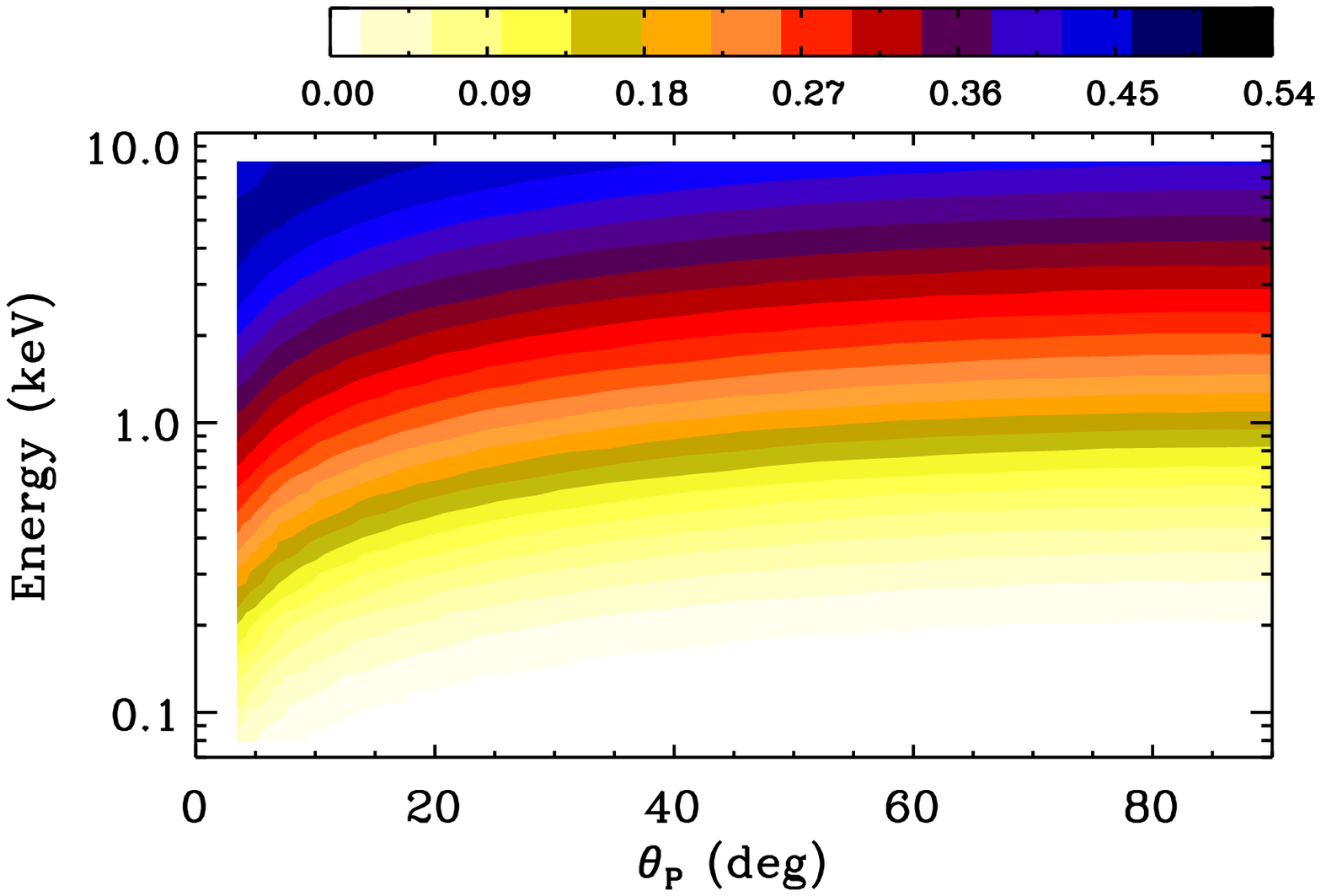} 
\caption{Photon-axion conversion probabilities for the magnetic field strength
  $B=10^{13}$~G and various
  combinations of the coupling strength, $g_a$, and axion mass, $m_a$.
  The probability is shown as
  a function of photon energy and angle $\theta_p$ between
  ${\bf B}$ and the direction of photon propagation.  In the
  {\em left} panels, the dependence on the coupling strength is explored
  for fixed axion mass $m_a=2\times 10^{-6}$~eV (from top to bottom:
  $g_a=10^{-8}, 10^{-9}, 10^{-10}$~GeV$^{-1}$).  The {\em right} panels show
  the dependence of the conversion probability on the axion mass
  for fixed coupling strength $g_a=10^{-8}$~GeV$^{-1}$ (from top to
  bottom: $m_a=7\times 10^{-7}, 6\times 10^{-6}, 1.7\times
  10^{-5}$~eV).  The dependence of the probability on $B$ is
  weak for the range considered here ($4\times 10^{12}-4\times
  10^{13}$~G), so we only display results at the intermediate field
  strength.
\label{fig:probs}}
\end{figure*}

The mixing strength can be measured in terms of the ratio between the off-diagonal
term and the difference between the diagonal terms in Eq.~(\ref{eq:evolution}); a useful
parametrization can be made in terms of the mixing angle $\theta_m$ defined by
\beq
\theta_m \;= \;\frac{1}{2}\arctan\left[\; \frac{2\,\Delta_M}{\Delta_a\; - \;\Delta_\parallel}\; \right]\;.
\label{eq:tetam}
\eeq
We begin our analysis by considering a region of space over which the magnetic
field can be approximated as homogeneous. Following
Raffelt \& Stodolsky (1988), the discussion is greatly simplified if we
define phases relative to the unmixed component, and
neglect a common phase.  The solution can be found by 
first performing a matrix rotation to an eigenstate basis where the
propagation matrix is diagonal, propagating the two eigenstates
independently, and then rotating back to the photon--axion basis.  
This yields an evolution equation for the mixing components: 

\beq \left[\begin{array}{c}
    A_O(z)\\
    a(z)\end{array}\right] \;= \;{\cal M}(z)
\left[\begin{array}{c}A_O(0) \\ 
                       a(0)                   
\end{array}\right],
\label{eq:evolution2}
\eeq

where
\beq
{\cal M}(z)= \left(\begin{array}{cc}\cos \theta_m & -\sin\theta_m\\ \sin\theta_m & \cos\theta_m\\ 
\end{array}\right) {\cal M_D}\left(\begin{array}{cc}\cos\theta_m  & \sin\theta_m\\ -\sin\theta_m & \cos\theta_m\end{array} \right)
\eeq
 and
 \beq
 {\cal M_D}=\left(\begin{array}{cc} e^{-i(\Delta'_\parallel-\Delta_\parallel)z} &0\\ 
0& e^{-i(\Delta'_a-\Delta_\parallel)z}  \end{array}\right)\,,
 \eeq
\beq
\Delta' = \frac{\Delta_\parallel\;+\; \Delta_a}{2}\; \pm\; 
\frac{\Delta_\parallel\;-\;\Delta_a}{2\,\cos{2\theta_m}}\,, 
\eeq
and the plus and minus sign indicates ${\Delta'}_\parallel$ and ${\Delta'}_a$, respectively.

The solution provided by Eq.~(\ref{eq:evolution2}) assumes that the field
is homogeneous.  The generalization for photon propagation within the
dipolar field around the NS, with $B\propto r^{-3}$, can be made
by assuming that the total evolution operator is the product of
evolution operators within spatial slices of width $\Delta z_j$ wherein
the $B$ field can be approximated as constant.

Within each shell, we use Eq.~(\ref{eq:evolution2}) to evolve the
photon and axion amplitudes from $A_O(z_j)$ and $a(z_j)$ in shell $j$
to $A_O (z_i) $ and $a(z_i)$, where $z_i=(z_j + \Delta z_j)$ is the
subsequent shell.  We tested the code for convergence and found that,
for $\Delta z_j\le 0.07$~km, the amplitudes at infinity converge for
the range of $B$, $m_a$, and $g_a$ of interest.

Fig.~\ref{fig:probs} shows the conversion probability for combinations of
the axion parameters $m_a,\; g_a$ in the soft X-ray band of interest for NS
observations.
The probability is plotted as a function of photon energy and propagation 
angle $\theta_p$.  There is a range in
the parameter space of axion mass, coupling
strength, and NS properties (such as surface magnetic field strength) where the
photon-axion conversion probability is non-negligible.
This implies that photon-axion conversion can leave
distinctive signatures in NS spectra, making them interesting sources for
constraining axion physics. 
In the next section, we perform a detailed analysis of such signatures.

\section{Thermal spectra, light curves, and polarization of neutron stars with
photon-axion conversion}

The observed spectrum of a NS is obtained by summing the local
emissivities over the emission region.  For simplicity, we consider
an emitting region that has a constant temperature and magnetic field strength, with the
field direction along the surface normal (e.g., a hot magnetic polar cap).
Such a region allows us to directly examine the spectral
dependencies of photon-axion conversion on these physical quantities.
It is often the case that the emission is dominated by a region 
smaller than the entire star surface (i.e., a hot spot).  The 
spectrum and lightcurves will then be functions of the viewing angle $\alpha$ 
between the observer and spot axes.
As we will show in \S~\ref{section:results}, phase
dependent methods add considerable diagnostic power for 
detecting the presence of axions.

The calculation of phase-dependent emission from an extended
region on the NS surface follows the formalism developed by Pechenick et al (1983),
with generalizations by Perna \& Gotthelf (2008; see also Bernardini et al.
2011, for a similar geometry). We define the time-dependent
rotational phase $\gamma(t)$ as the azimuthal angle subtended by the
magnetic dipole vector {\boldmath{$\mu$}} around the axis of rotation. It is
related to the modulus of the NS angular velocity, $\Omega(t)$, by $\gamma(t)=\Omega(t) t$. 
We choose the coordinate system so that the observer is located along
the ${z}$ axis; the inclination angle of the rotation axis, {\boldmath$\hat{\Omega}$}, 
with respect to the line of sight is denoted by $\alpha_R$, while the
angle between the magnetic dipole vector and the rotation angle is denoted
by $\alpha_M$. Thus, the angle between {\boldmath{$\mu$}} and the line of sight is
given by 

\begin{eqnarray}
\label{eq:alpha}
\cos\alpha(t) = \cos\alpha_R\cos\alpha_M+\sin\alpha_R\sin\alpha_M\cos\gamma(t).
\end{eqnarray}

\noindent The emission region is assumed to be circular, with opening angle
$\beta$, and is centered around the magnetic dipole axis.  Hence,
$\alpha(t)$ also indicates the time-dependent angular separation
between the spot axis and the line of sight as the star rotates.  We
describe points on the NS surface by means of the polar angle
$\theta$, and the azimuthal angle, $\varphi$, in spherical polar
coordinates (see Fig.~1 in Perna \& Gotthelf 2008 for a visual representation
of this geometry).

The emission region is restricted to $\theta\le\beta$ for $\alpha=0$, while for $\alpha\ne 0$ it is
identified by the conditions

\beq
   \left\{
  \begin{array}{ll}
    \alpha-\beta\le\theta\le\alpha+\beta \\
      2\pi-\varphi_\star\le\varphi\le\varphi_\star\;\;\;\; \;\;\;\;\;\;\;\;{\rm if}\;\;\;\beta\le\alpha\\
  \end{array}\right.\;
\label{eq:con2}
\eeq

\noindent with
\beq
\varphi_\star=\arccos\left[\frac{\cos\beta-\cos\alpha\cos\theta}{\sin\alpha\sin\theta}\right]\;,
\label{eq:phip}
\eeq

\noindent and by the condition

\beq
\theta\le\theta_*(\alpha,\varphi,\beta),\;\;\;\;\;\;\;\;\;\;\;\;
{if}\;\;\; \beta > \alpha\; .
\label{eq:con3}
\eeq

\noindent In the latter case, the outer boundary of the spot, 
$\theta_{\star}(\alpha,\varphi,\beta)$, must be determined numerically from the expression

\beq
\cos\beta  =  \sin\alpha\sin\theta_{\star}\cos\varphi+\cos\alpha\cos\theta_{\star}\,.
\label{eq:numeric}
\eeq

Due to the intense gravitational field of the NS, photons
emitted at the NS surface are substantially deflected as they travel
to the observer. A photon emitted from colatitude $\theta$ will reach 
the observer (at infinity) if emitted at an angle $\delta$ with respect to the
surface normal; the relation between the two angles is given by
the ray-tracing function 

\begin{eqnarray}
\theta(\delta) = \int_0^{R_s/2R}\,du\;\, x\,\left[\left(1-{R_s\over R}\right)\left({R_s\over 2R}\right)^2-
	(1-2u)u^2 x^2\right]^{-1/2}\,, 
\label{eq:teta}
\end{eqnarray} 
\noindent where $x\equiv\sin\delta$, $R$ and $M$ are the NS radius and mass,
respectively,
and $R_s\equiv 2GM/c^2$ is the Schwarzchild radius.

The spectrum at the observer is obtained, as a function of
the viewing angle, $\alpha(\gamma)$, by integrating the local emission
over the observable surface; this procedure yields the 
flux seen by an observer at a distance $D\gg R$.  Accounting for the
gravitational redshift of the radiation, this integral takes the form (Page 1995 and
generalizations by Pavlov \& Zavlin 2000, Heyl et al. 2003):
\begin{eqnarray}
F_j(E_\infty,\alpha) \;=\;\frac{2 \pi}{c\,h^3}\;\frac{R_\infty^2}{D^2}\;E_\infty^2
\int_0^1 2xdx\nonumber \\ 
 \times \int_0^{2\pi}\; \frac{d\phi}{2\pi}\; 
I_j\,\left(\theta,\,\phi,\,E_\infty e^{-\Lambda_s}\right)\;,
\label{eq:flux}
\end{eqnarray} 
in units of photons~cm$^{-2}$~s$^{-1}$~keV$^{-1}$.  
\noindent {In the equation above, the index $j$ represents any of the Stokes parameter (U, V, Q, I)}, and 
$E_\infty$ is the energy as seen by the
distant observer.  The energy emitted at the stellar surface is given by $E_\infty e^{-\Lambda_s}$, with 

\beq
e^{\Lambda_s}\equiv\sqrt{1-{\frac{R_s}{R}}}, 
\eeq 
The local specific intensity
$I_j$ is equal to zero outside of the
boundaries for $\theta$ and $\varphi$ defined by equations (\ref{eq:con2})-(\ref{eq:con3}).

The observed flux depends on the geometric angles $\alpha_R$ and
$\alpha_M$ through the angle
$\alpha$ in Eq.~(\ref{eq:alpha}).  In the following, we consider an
orthogonal rotator, for which $\alpha_R=\alpha_M=90^{0}$.  This implies that
$\alpha(t)=\gamma(t)$.  Given that the emission region is centered
around the dipole axis, the angle $\alpha$ will be identified
with the angle $\theta_p$ between the photon propagation and 
magnetic field directions (see \S\ref{section:prob}).

If photon-axion conversion occurs with probability $P(E_\infty,\alpha)$,
the phase-resolved spectrum is given by

\beq
F(E_\infty,\alpha) \;= F_{\rm X}(E_\infty,\alpha)+ [1 - P(E_\infty,\alpha)]\;
F_{\rm O}(E_\infty,\alpha).
\label{eq:spectrum}
\eeq

\noindent For a fixed energy (or integrated over an energy band), $F(E_\infty,\alpha)$ 
(or $\int_{E_1}^{E_2}dE_\infty F[E_\infty,\alpha]$) yields the light curve as a function of the
viewing angle $\alpha$. 
The phase averaged spectrum is readily obtained from

\beq
F_{\rm ave}(E_\infty)=
\frac{1}{2\pi}\int_0^{2\pi}d\alpha \; F(E_\infty,\alpha)\;.
\label{eq:Fave} 
\eeq

\noindent Finally, the linear polarization is computed as
\beq
\Pi(E_\infty,\alpha)\;=\; \frac{[1 - P(E_\infty,\alpha)]F_{\rm O}(E_\infty,\alpha) \; - \; 
F_{\rm X}(E_\infty,\alpha)}
{[1 - P(E_\infty,\alpha)]F_{\rm O}(E_\infty,\alpha) \; + \; F_{\rm X}(E_\infty,\alpha)}\;\;.
\label{eq:pol}
\eeq

\noindent To simplify our notation, we will omit the subscript $`\infty'$ from the observed energy.  
In the discussion that follows, all quoted energies are redshifted energies for a
NS of mass $M=1.4M_\odot$ and radius $R=10$~km.

Before concluding this section, we need to remark that
Eq.~(\ref{eq:spectrum}) and Eq.(\ref{eq:pol}) assume that the O and X
modes remain decoupled once they have emerged from the atmosphere of
the NS.  The evolution of the modes in the changing dipole magnetic
field of the star was computed in papers by Heyl \& Shaviv (2000,
2002), Heyl, Shaviv \& Lloyd (2003), Lai \& Ho (2003) and van
Adelsberg \& Lai (2006).  All of the above studies found that, in the
birefringent, magnetized vacuum near the star surface, the photon
modes are decoupled.  In this region, the mode eigenvectors evolve
adiabatically along the changing direction of the magnetic field.
Further from the star, the mode evolution continues to be adiabatic,
up to the "polarization limiting radius", ${R_{pl}}$, which is defined
as the distance at which the polarization modes mix (and are frozen
thereafter).  { Heyl, Shaviv \& Lloyd (2003) derive the following
  expression for $R_{pl}$\footnote{A similar expression for the
    polarization limiting radius is derived in van Adelsberg \& Lai
    (2006): $R_{pl} =
    7.2\left({E_1\,B^2_{13}[\Omega/(2\pi)]^{-1}\sin^2\alpha}\right)^{1/6}
    \,10^7 \,{\rm cm}$.  The scaling of $R_{pl}$ in this equation
    differs slightly from that in Eq.~(\ref{eq:Rpl}) due to different adiabatic
    conditions applied by the two sets of authors.  Heyl et al. (2003)
    constrain the change in the relative difference in the mode
    indices of refraction compared to the magnitude of that
    difference [see their Eq.~(9)], while van Adelsberg \& Lai (2006)
    consider the distance at which the mode evolution equations have
    equal diagonal and off-diagonal terms [see their Eq.~(62)].  
    This results in slightly 
    different magnitudes and scaling relations; for slowly rotating 
    NSs, the precise numerical value of $R_{pl}$ does not change 
    the value of the observables, and both approaches result in the 
    same physics.}  
\beq R_{pl}\;\approx \;
  3.6\times 10^7\;B_{13}^{2/5}\;E_1^{1/5}\;(\sin\alpha)^{2/5}\;\;{\rm
    cm}\,,
\label{eq:Rpl}  
\eeq where $E_1\equiv E/(1\,{\rm keV})$ and $B_{13}\equiv
B/(10^{13}\,{\rm G})$}.
For radii larger than $R_{pl}$, the mode
eigenvectors are fixed, and the polarization no longer evolves (i.e.,
it is the observed polarization).  The polarization modes do couple
near the polarization limiting radius.  The effect of this coupling
was explicitly computed, for example, in van Adelsberg \& Lai (2006).
However, except for very rapidly rotating stars (i.e., with
frequencies of tens of milliseconds) the coupling results in only a
very small quantitative change.  For example, for a star with period
$\sim 0.1$~sec ($B\sim 10^{12}$~G, $E\sim 1$~keV), the mode
intensities $I_{\rm O}$ and $I_{\rm X}$ ($\propto \left|A_{\rm
  O}\right|^2$, $\left|A_{\rm X}\right|^2$) couple at the level of
$\la 4\%$, and the amount of mixing further decreases at longer
periods.  Therefore, for the purpose of our study and the
observational tests we propose (requiring NSs with relatively long
periods in order to perform time-resolved spectroscopy), the formalism
that we use provides a very good description of the observed flux and
the linear polarization fraction computed here (see also Lai \& Heyl
2006). {We further note that, if the radius of the emission region
  is much smaller than the polarization limiting radius, then summing
  over the Stokes parameters at the detector is well approximated by
  summing over a mode (X or O) at the NS surface (Heyl et
  al. 2003). Since our emission region (a hot spot) is always much
  smaller than $R_{pl}$ for the $B$ field strengths and energies under
  consideration, we use this approximation in our calculations.}

\begin{figure*}[ht]
\includegraphics[scale=0.9,angle=0]{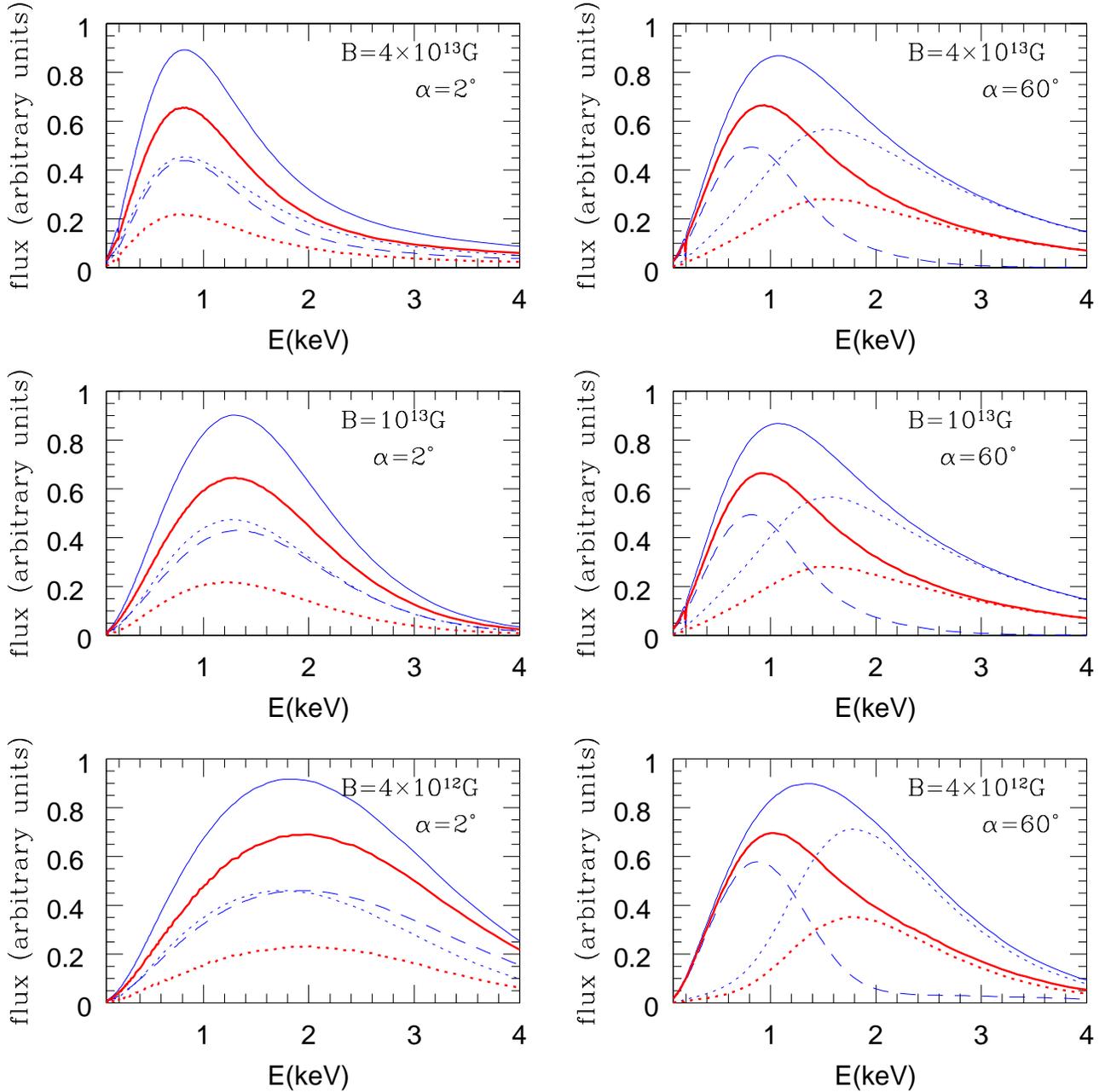}
\caption{Comparison between phase-resolved spectra of NSs with {\em
    (thick/red lines)} and without {\em (thin/blue lines)} photon-axion conversion. In each
  case, the dotted lines show the flux contribution from the O mode,
  while the dashed lines display the contribution from the X mode,
  which is unaffected by conversion. The solid lines show the
  total spectra. In all panels, the effective temperature of the star
  is $T_s=5\times 10^6$~K and the axion parameters are
  $m_a=2\times 10^{-6}$~eV and $g_a=10^{-8}$~GeV$^{-1}$, while the $B$-field
  strength has three values from top to bottom.  The
  spectra are shown for two angles $\alpha$ between the
  observer and the dipole axis.
\label{fig:spectra}}
\end{figure*}

\section{Results: theoretical predictions and observational tests}
\label{section:results}

\subsection{Theoretical predictions}
\label{subsection:theory}

We present the results of our calculations of
NS spectra, light curves, and polarization signals, with and without photon-axion
conversion.  We explore the dependencies on the stellar magnetic field
strength and effective temperature, with particular
emphasis on the magnetic field since it has a
stronger influence than the temperature on the functional form of the
modal energy dependence (see Figs.~\ref{fig:fluxesB} and
\ref{fig:fluxesT}).  We consider emission from a circular region,
10~deg in angular size, that is centered on the magnetic pole of the
star. Such hot spot emission is common in NSs,
and exhibits a strong dependence on phase.  Given the significant angle
dependence of the O and X mode intensities (see
Figs.~\ref{fig:fluxesB} and \ref{fig:fluxesT}), the study of emission from
a finite region allows us to extract more information on the effects that
photon-axion conversion has on the observable properties of NSs.

\begin{figure}[ht]
\includegraphics[scale=0.45,angle=0]{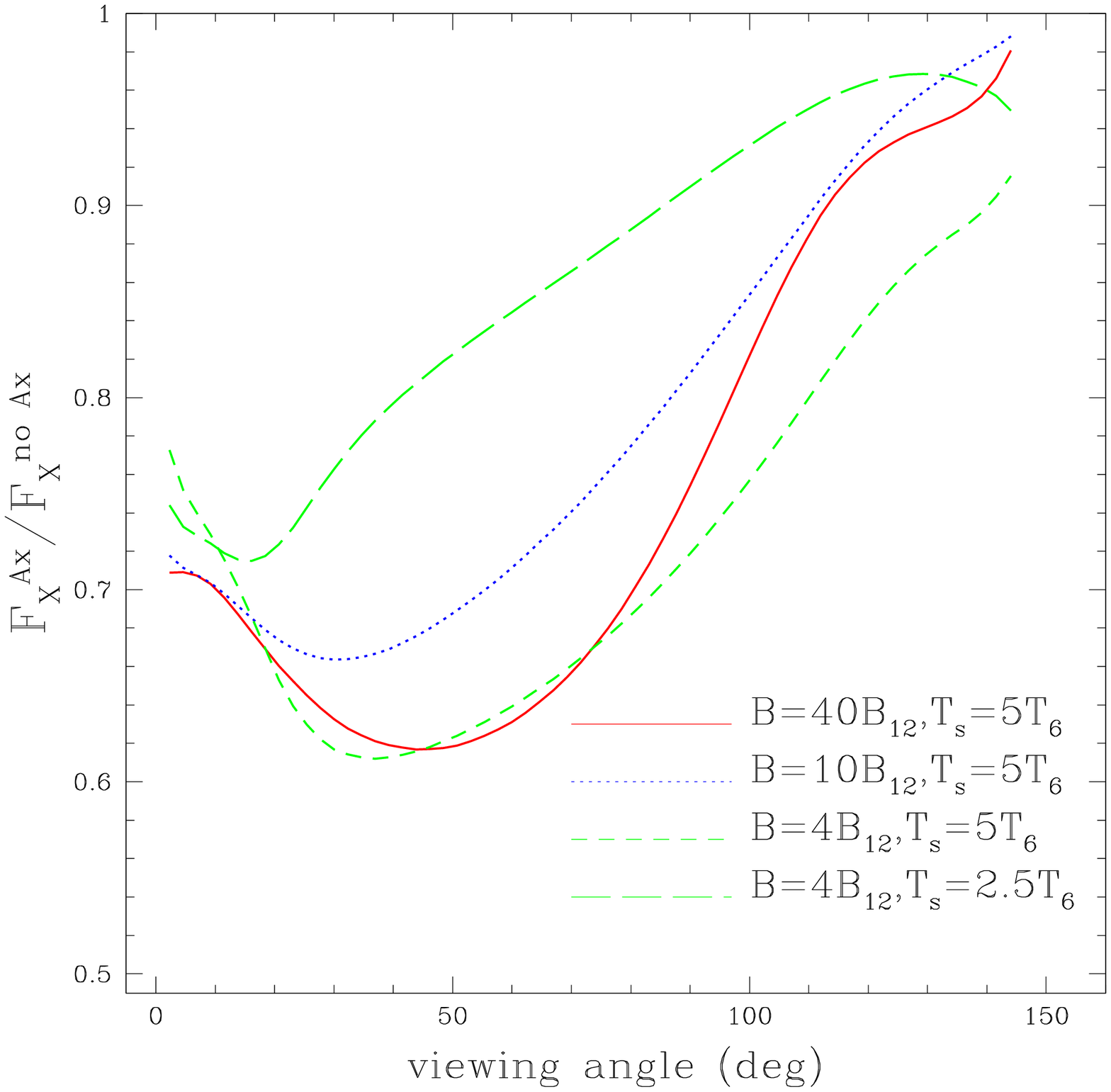}
\caption{Ratio of X-ray flux (in the [0.08-8] keV band) 
  with axion conversion to that 
  without axion conversion.  Axion parameters have the same values as in
  Fig.~\ref{fig:spectra}, while the NS temperature and magnetic field are
  indicated for each curve, with the $B_{12}\equiv
  10^{12}$~G and $T_6=10^6$~K.
\label{fig:flratio}}
\end{figure}

As shown in Fig.\ref{fig:probs}, the conversion
probability is a strong function of the axion mass, $m_a$,
and the coupling constant, $g_a$.  For the thermal spectra of
NSs to be affected by photon-axion conversion,
the probability of this process must be non-negligible in the soft X-ray
band, where the NS thermal spectrum peaks.  Therefore, we
perform a detailed analysis of the consequences of photon-axion conversion on
NS spectral properties for a particular set of axion parameters;
these parameters are chosen to yield a substantial conversion probability in the 
soft X-ray band.  We choose the values $m_a=2\times 10^{-6}$~eV, and
$g_a=10^{-8}$~GeV$^{-1}$; the study of this specific case will uncover
the main signatures of axion conversion.  Once these are identified, we 
explore a wider area of $m_a-g_a$ parameter space to
identify the main region that can be probed by means of NS
observations in the X-ray band.

Fig.~\ref{fig:spectra} shows phase-resolved spectra for three magnetic
field strengths, $B=4\times 10^{13}$, $10^{13}$, and $4\times
10^{12}$~G.  For each value of $B$, we plot the spectra at two phases,
coinciding with the angles between the line of sight and the magnetic
field axis for the chosen geometry.  We show angles $\alpha=2$~deg and
a typical large viewing angle $\alpha=60$~deg.  The former angle is
essentially a 'phase-on' spectrum (i.e. a spectrum observed in
correspondence to the maximum of the pulsation); we do not plot the
results for $\alpha=0^\circ$ since the conversion probability in that
case is zero.  For each combination of viewing angle and magnetic
field strength, the figure shows the fluxes from the O and X-modes
separately, as well as the total composite spectra. Results are
plotted with and without photon-axion conversion.

The relative fraction of O and X mode photons, as a function of energy
and angle, plays a fundamental role in determining the effects of
axions on phase-dependent spectra and light curves.  Before discussing
the spectral effects of axions in further detail, we comment on
the energy and angular dependence of the two photon modes.
Although only the O-mode photons are affected by axion conversion,
the properties of the X-mode photons also matter
for our purposes because of vacuum polarization-induced mode
conversion.  The X-mode has a complex beam pattern, consisting of the sum of a
narrow ``pencil'' beam and a broad ``fan'' beam.
The width of the pencil beam scales as $\sim (E/E_{Be})^{1/2}$, 
decreasing with magnetic field strength (Pavlov et al. 1994).
Therefore, the modal intensities vary from being approximately 
equal to significantly different between widely separated emission angles.

For a phase-on spectrum, the intensities of the O and X modes
peak at roughly similar energies.  Thus, the presence of photon-axion
conversion results in a dimming of the spectrum without
significant spectral distortion.  The situation is different at
larger viewing angles, as evident in the $\alpha=60$~deg panels of the
Fig.~\ref{fig:spectra}.  In these cases, the X mode dominates the spectrum 
at (redshifted) energies $\la 1.4$~keV, while the O mode dominates 
at higher energies; thus, an inversion of the dominant spectral mode occurs.  
Suppression of the O mode by photon-axion conversion
therefore produces not only a dimming of the spectrum but also a {\em spectral
distortion}, suppressing the high-energy tail. 

Since axion conversion occurs at larger distances from the star
than mode conversion due to the vacuum resonance, only 
O-mode photons are affected as they emerge from the star atmosphere.  This is
shown in Fig.~\ref{fig:spectra}, where the X-mode photons are
untouched by photon conversion, while the O-mode intensities are
noticeably suppressed.  The overall effect on the spectrum is that of
flux suppression. However, since the O/X flux ratio, as well as the
conversion probability, has an energy dependence which varies with
viewing angle, the shape of the spectrum is also affected,
causing a shift of the energy peak. While the shift is very small at phase-on
viewing angles (since the O-modes and X-modes have a similar
energy-dependent intensity for $\alpha\sim 0$~deg), it is very
pronounced at wider viewing angles, where the O-mode largely
dominates the intensity at high energies. In this case, enhanced
suppression of high energy photons produces an effective shift
of the peak of the spectrum towards lower energies.

\begin{figure*}[t]
\includegraphics[scale=0.38,angle=0]{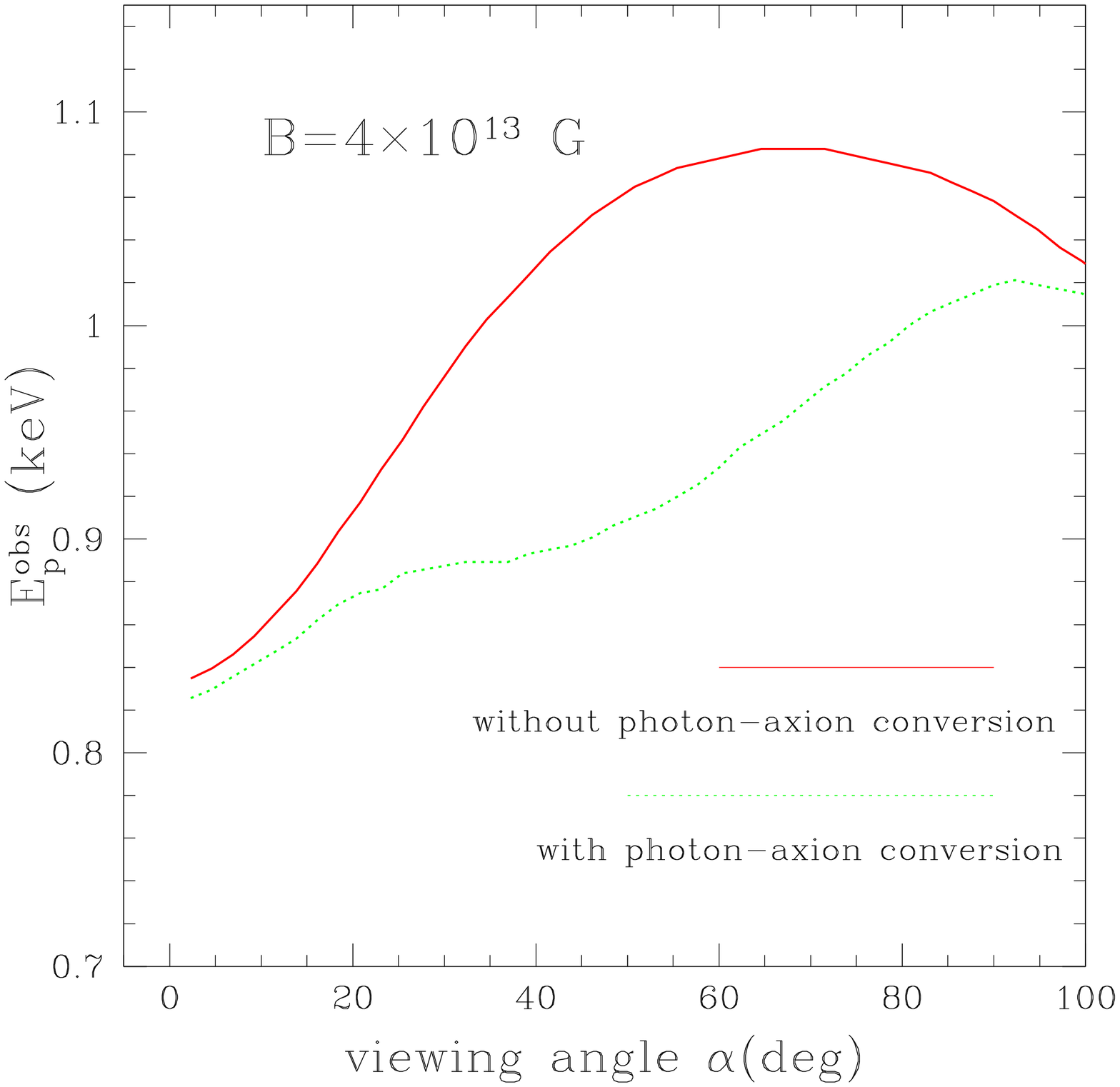}
\includegraphics[scale=0.38,angle=0]{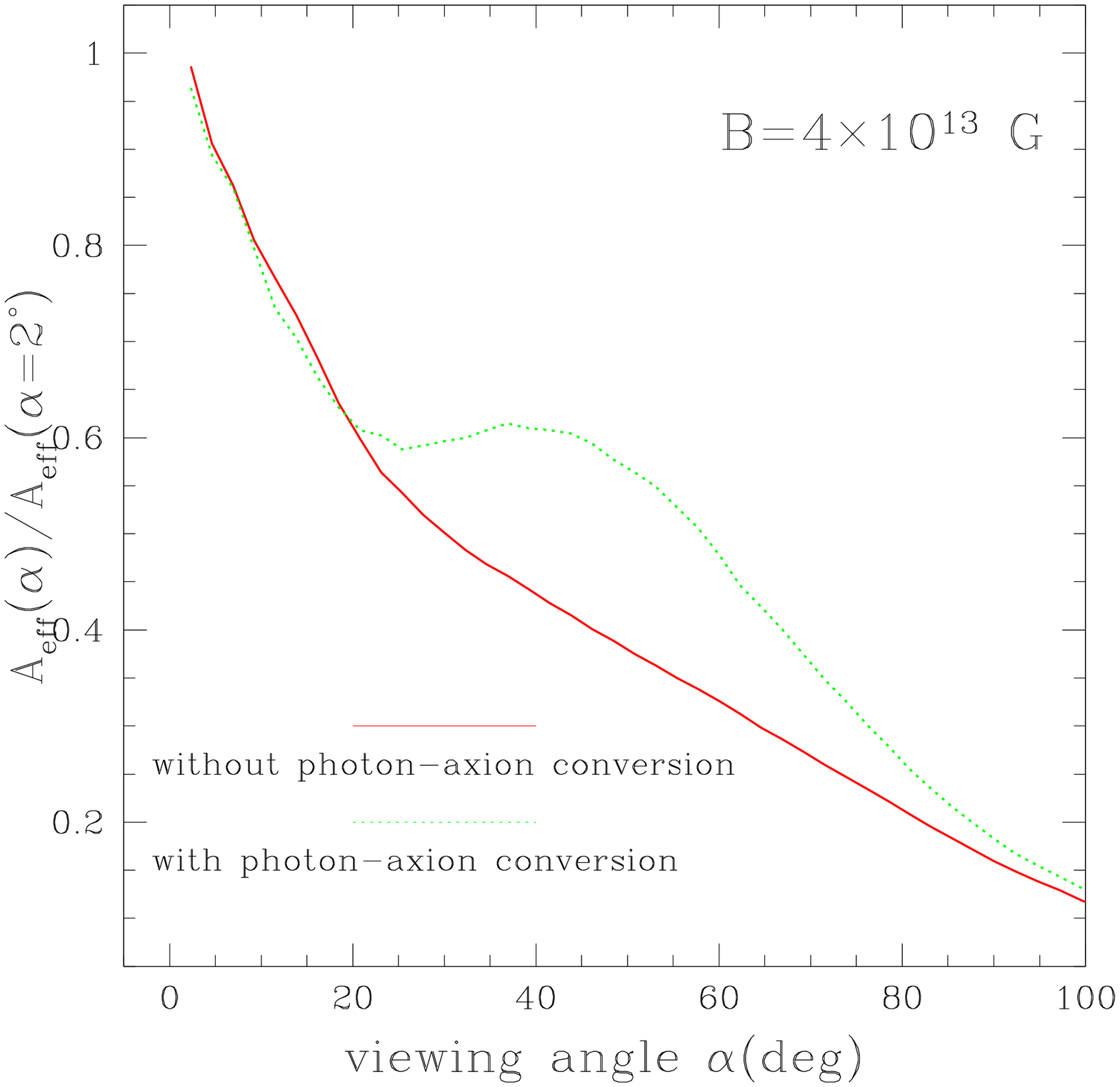}\\
\includegraphics[scale=0.38,angle=0]{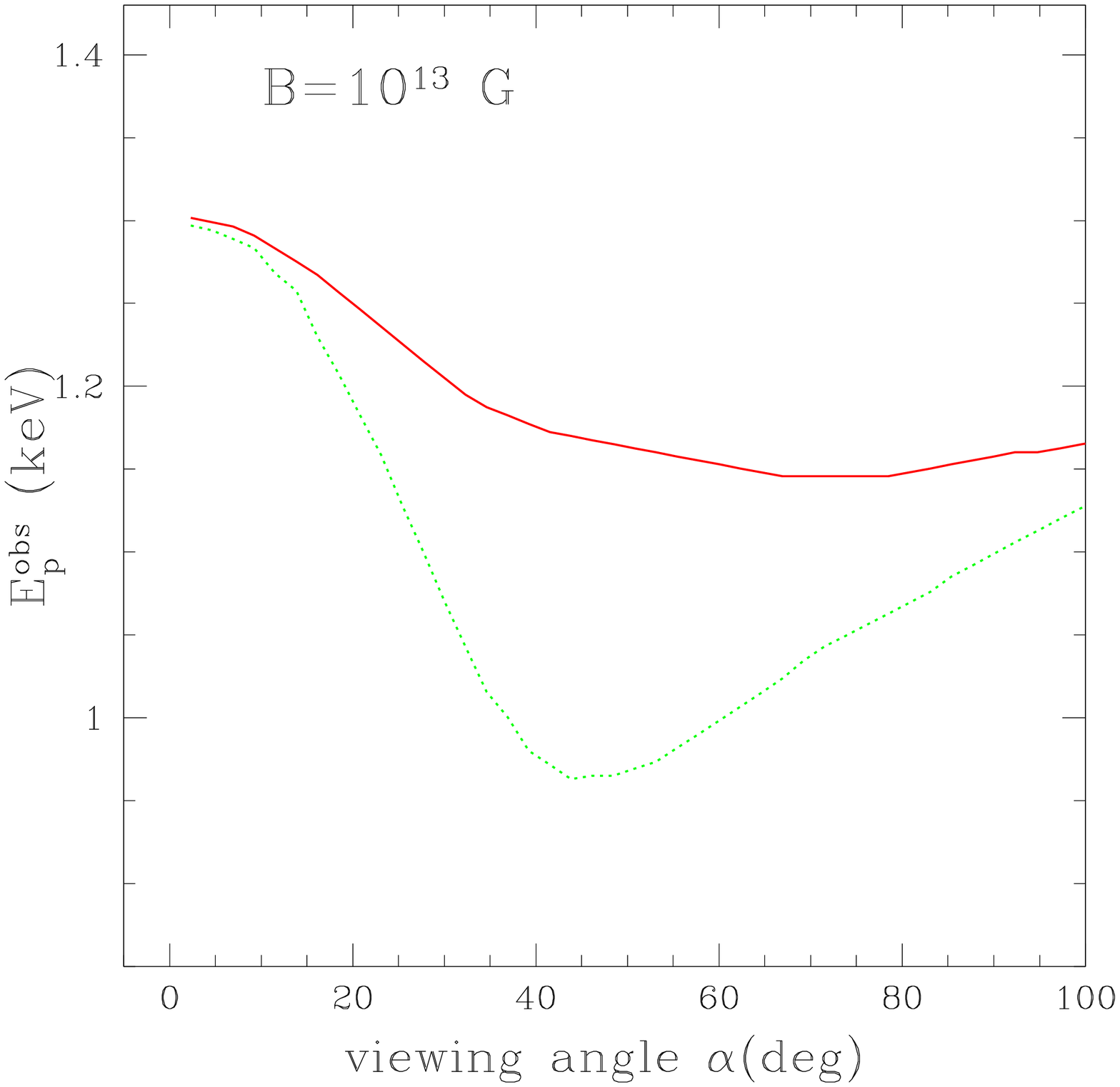}
\includegraphics[scale=0.38,angle=0]{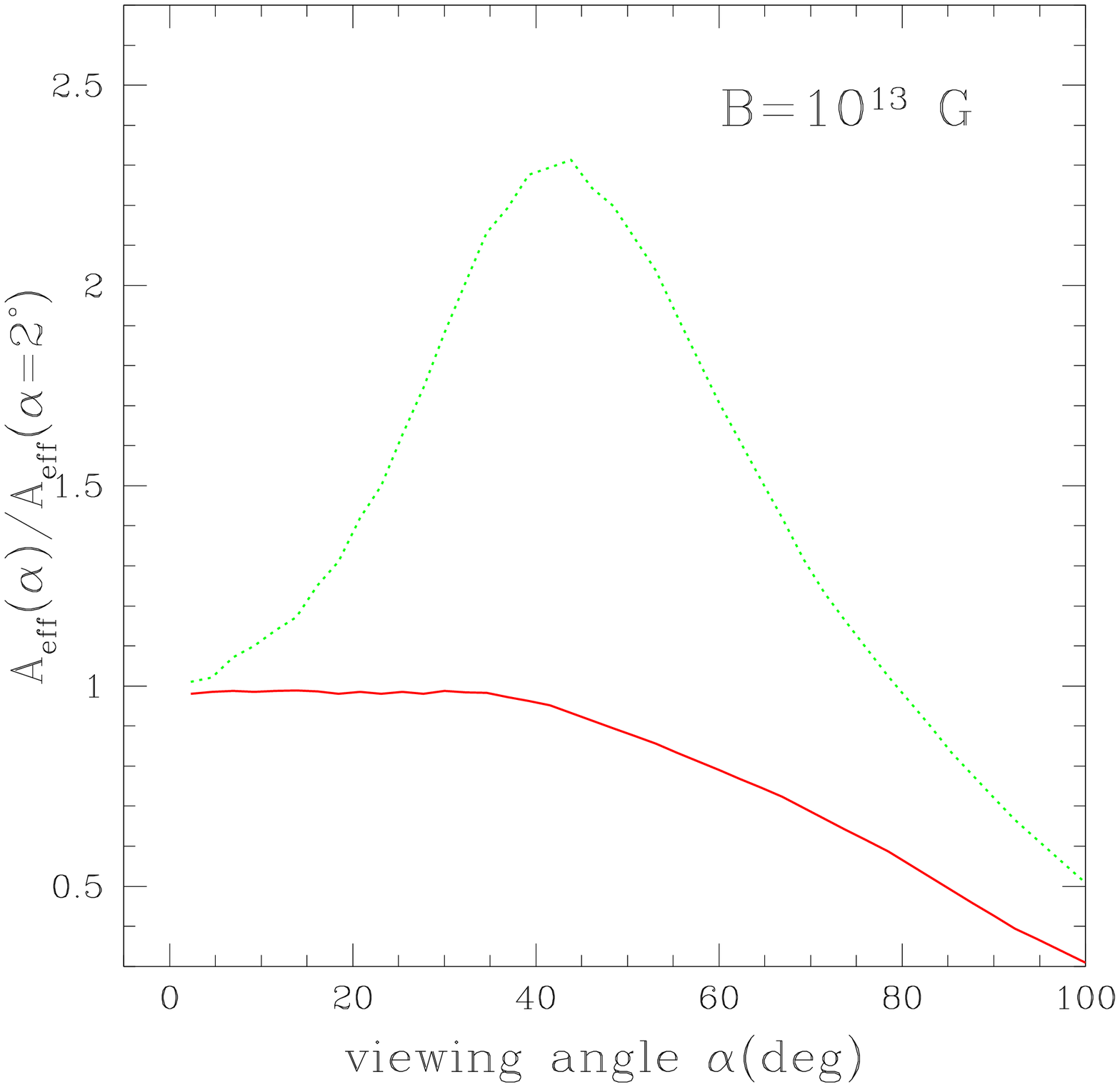}\\
\includegraphics[scale=0.38,angle=0]{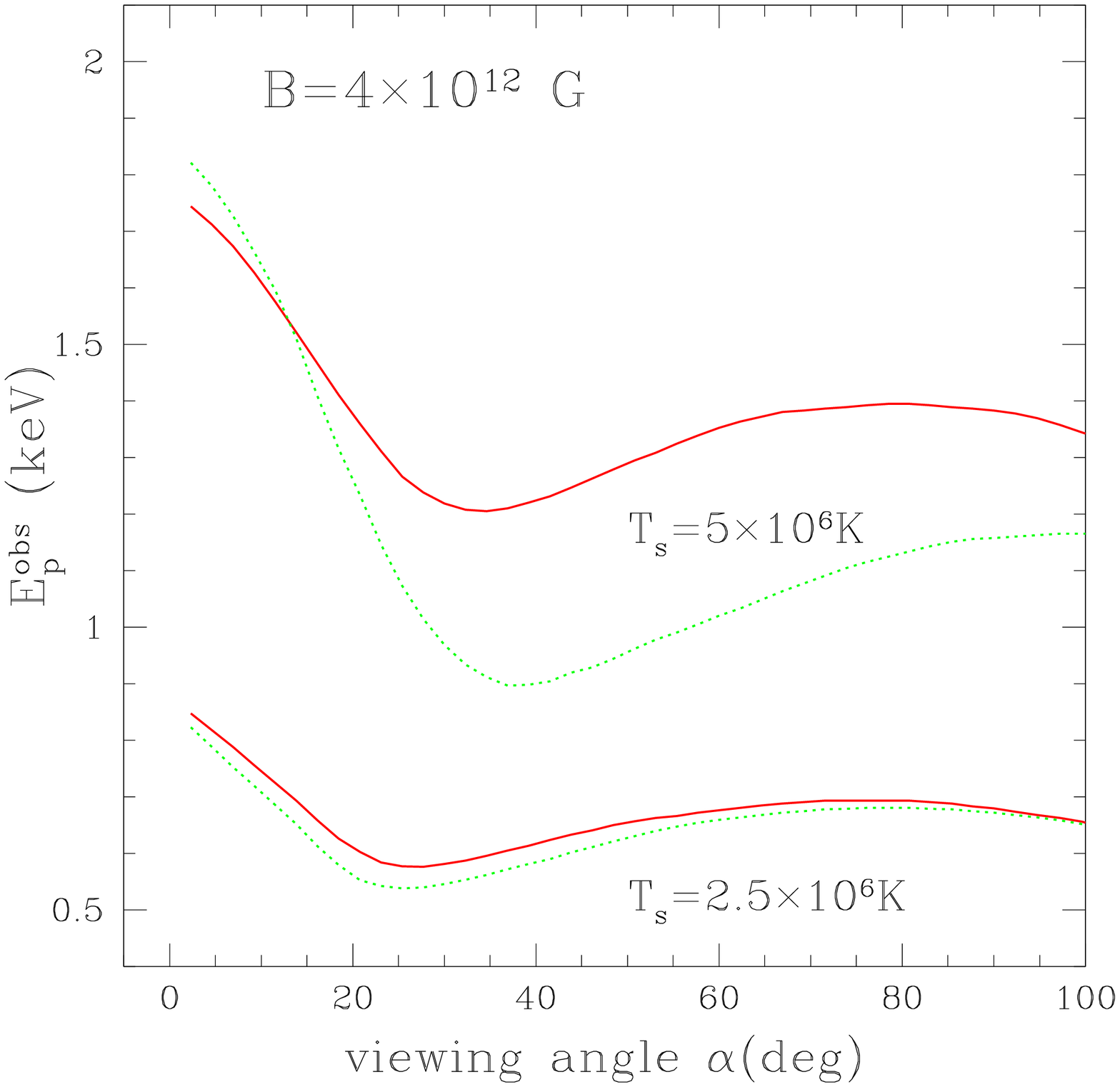}
\includegraphics[scale=0.38,angle=0]{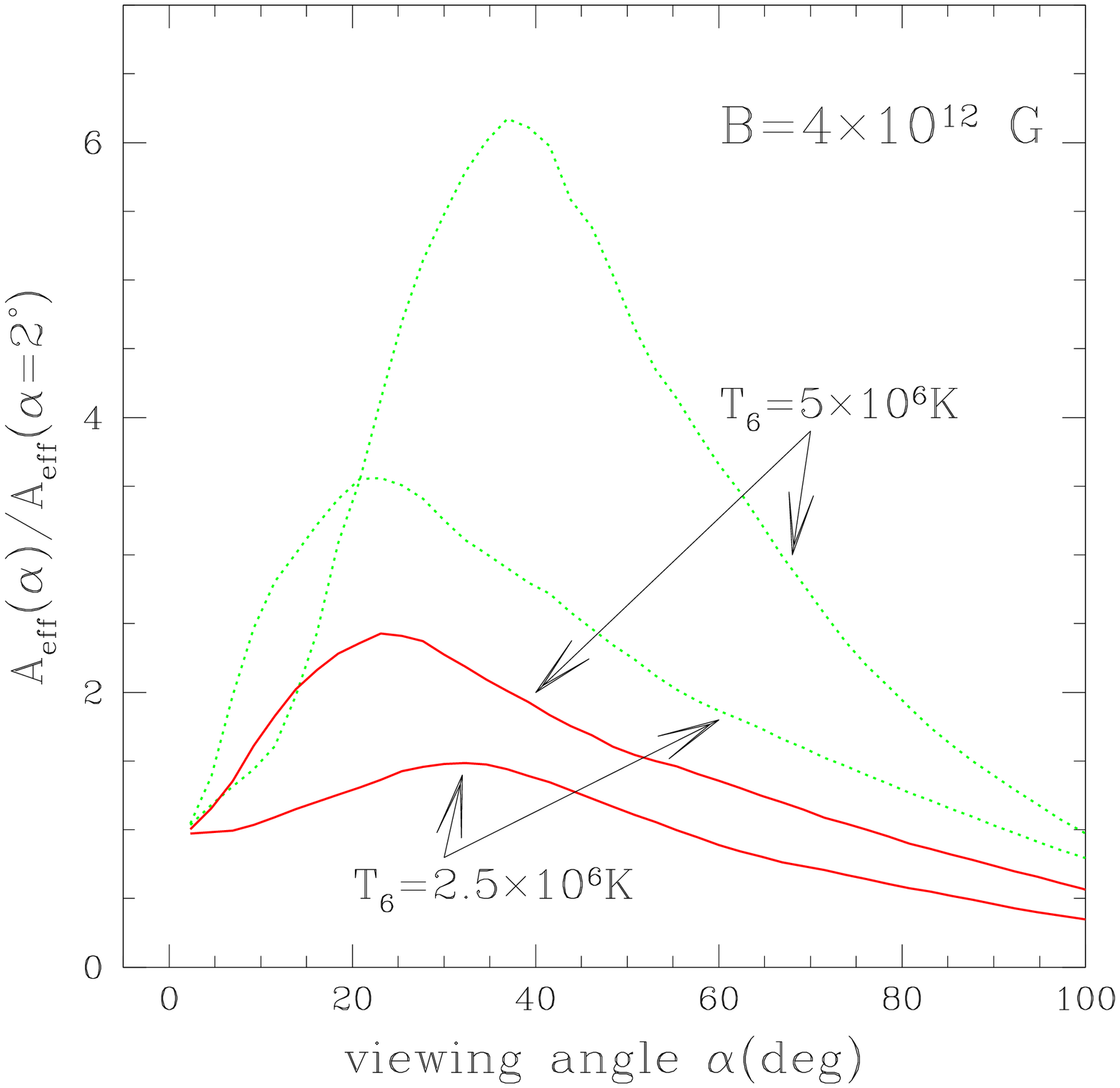}
\caption{{\em Left:} Peak energy of the observed spectrum as a function
of viewing angle $\alpha$ (i.e., phase of the NS). The presence of photon-axion
conversion changes and enhances the variation in peak energy as the star rotates.  
{\em Right:} Corresponding phase-resolved effective area that an observer would infer from spectral fitting.
Axion parameters have the same values as in Fig.~\ref{fig:spectra}. 
\label{fig:Aeff}}
\end{figure*}

Figures~\ref{fig:flratio} and \ref{fig:Aeff} quantify the phase-dependent
spectral effects discussed above.  For the same values of magnetic
fields and axion parameters considered in Fig.~\ref{fig:spectra},
Fig.~\ref{fig:flratio} shows the ratio of the X-ray flux in the
0.8-10~keV band photon-axion conversion to that without conversion.  As expected,
at all phases, the ratio $F_{X}^{\mathrm{Ax}}/F_{X}^{\mathrm{no Ax}}$
is smaller than
1. This is because the contribution to the flux from the O-mode
photons is non-negligible at all viewing angles with respect to the
magnetic field axis. The precise value of the flux ratio, however,
shows a significant dependence on the viewing angle, with a mimimum around
$\alpha\sim 40-50$~deg.  This is due to the 
O-mode flux, which is a large fraction of the total flux at $E\ga 1$~keV, for viewing angles in
the $30-70$~deg range.

\begin{figure*}[ht]
\includegraphics[scale=0.6,angle=0]{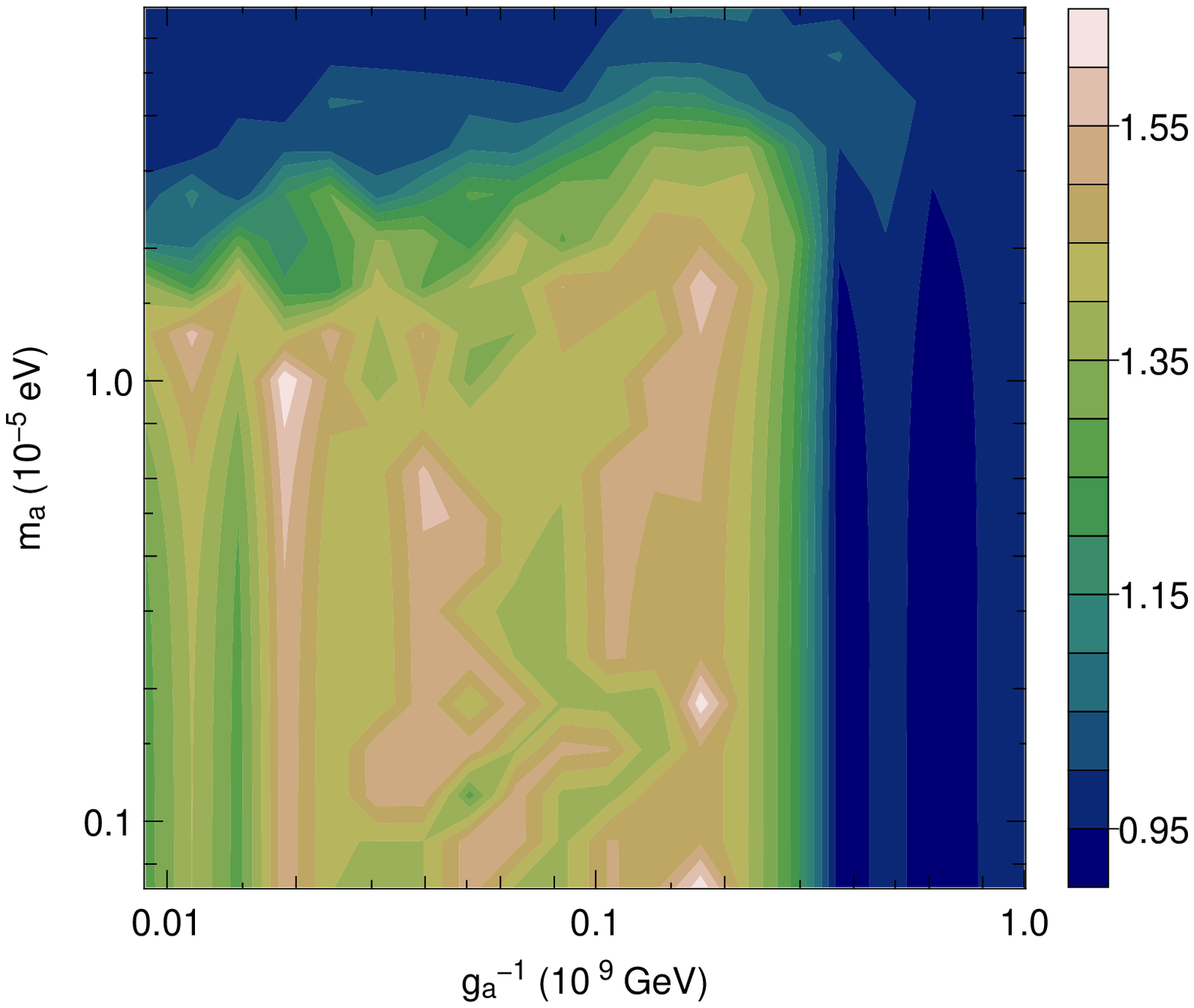}
\includegraphics[scale=0.6,angle=0]{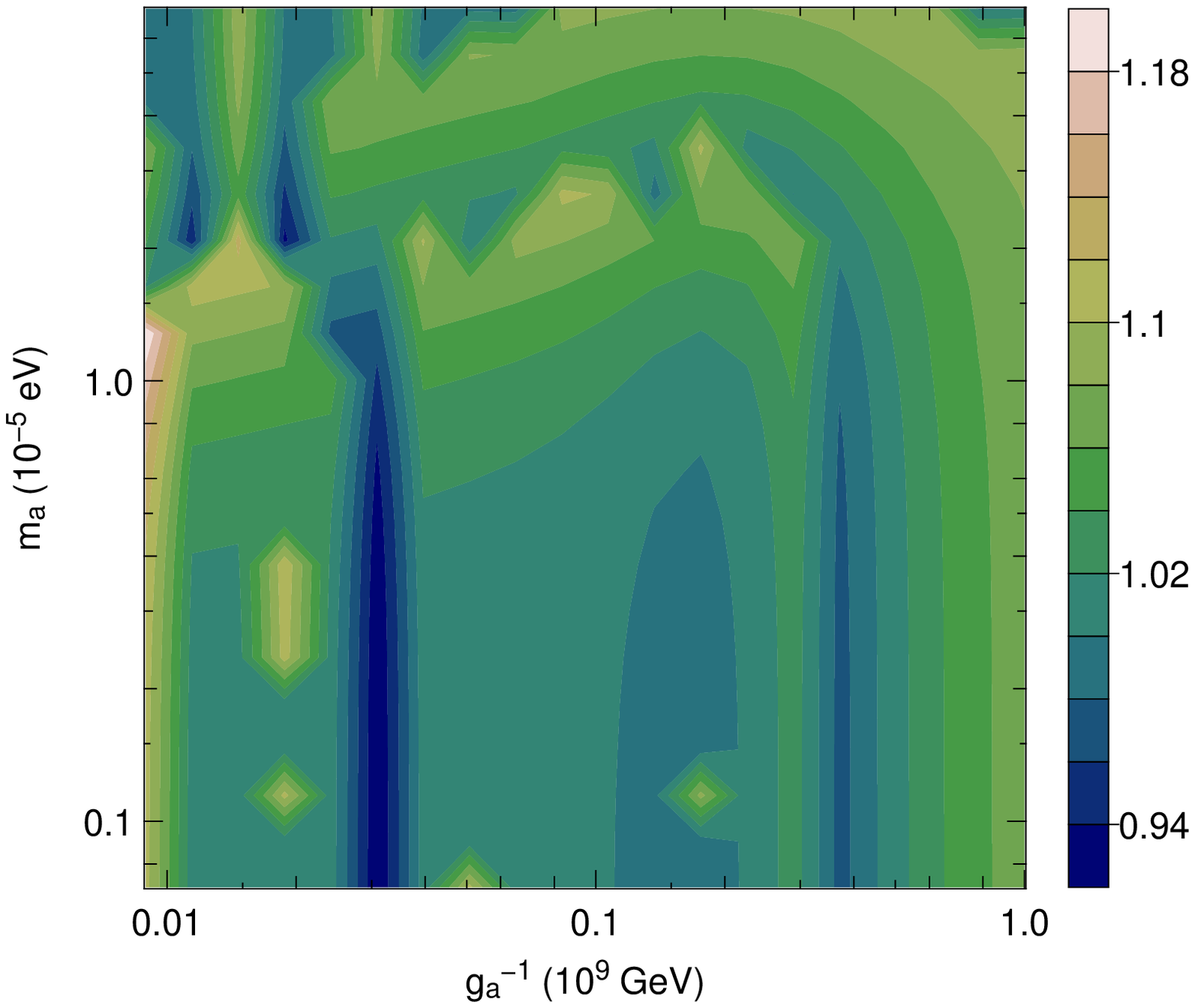}\\
\includegraphics[scale=0.6,angle=0]{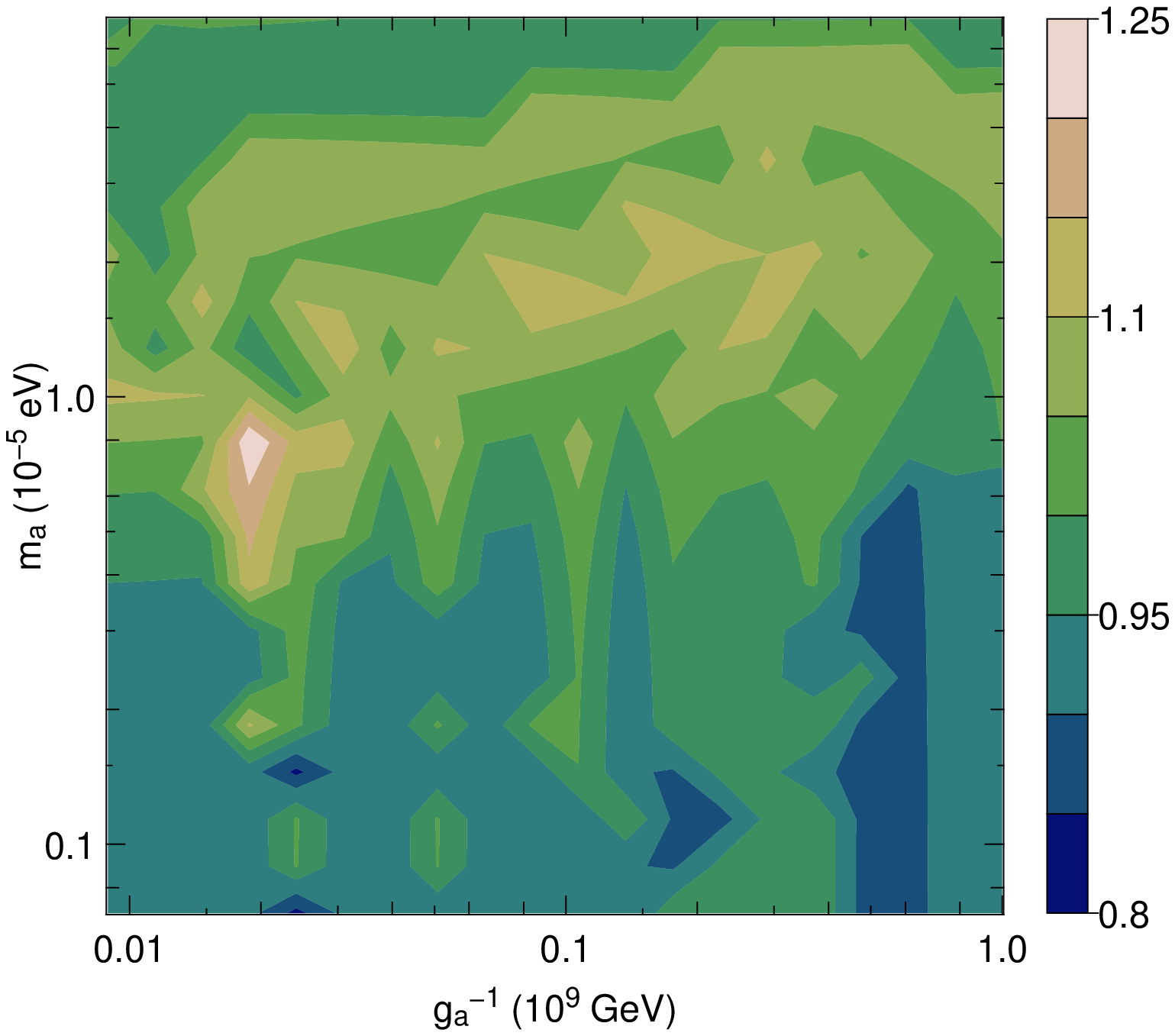}
\caption{Ratio between the observed effective area with
  photon axion conversion and without photon-axion
  conversion for a range of axion parameters $m_a$ and $g_a$. The NS
  magnetic field and temperature are $B=4\times 10^{12}$~G
  and $T=5\times 10^6$~K {\em (top left panel)}, $B=4\times 10^{13}$~G
  and $T=5\times 10^6$~K {\em (bottom left panel)}, and $B=4\times
  10^{12}$~G and $T=2.5\times 10^6$~K {\em (top right panel)}, respectively.
\label{fig:2D}}
\end{figure*}

The left panels of Fig.~\ref{fig:Aeff} show the (redshifted) energy
where the spectrum peaks, $E_{\rm peak}$, as a function of viewing
angle.  We consider only angles $\alpha\le 100$~deg since the flux from
larger angles, which is still visible due to relativistic
light deflection, is too low for phase-resolved spectroscopy.
We begin by examining the behavior of $E_{\rm peak}$ without photon-axion
conversion. At the magnetic field strengths for the models in this paper,
the spectral peak of the total (X and O mode) specific intensity follows
the beam pattern of the X mode: a narrow pencil beam at small angles
and a broad fan beam at large angles.
This is because conversion between the X and O modes occurs above the
photosphere of both modes and thus does not change the total intensity.
The observed variation of $E_{\rm peak}$ with viewing angle deviates somewhat from the 
emission beam pattern due to the narrowness of the pencil beam (a few deg in width
around $\delta=0^\circ$), the gravitational deflection of photon trajectories,  
and the finite size of the emitting region, which smears out features from nearby emitting
points\footnote{To be able to clearly observe the very narrow pencil beam, both the 
size of the emission region and the viewing angle would need to be smaller than (or comparable 
to) the width of the beam.}. 

When axion conversion is taken into account for $B = 4\times 10^{12} - 4\times 10^{13}$~G, 
a common feature is a shift of the peak energy to lower
values at all viewing angles except for those close to phase-on and
edge-on.
For $\alpha\approx 0$~deg, the intensity of the O mode photons peaks at
similar or slightly lower energies than that of the X-mode photons
(see Fig.~\ref{fig:spectra}); hence, suppression of O mode photons
primarily affects the amplitude of the spectral intensity, as opposed to the
location of the spectral peak.  In fact, because
the O-mode intensity peaks at lower energies, especially at
lower fields, at small angles there is actually a shift of the peak towards
higher energies after photon-axion conversion.  

Another notable signature in the variation of 
$E_{\rm peak}(\alpha)$ is a strong decrease (or flattening in the
higher $B$-field case) for viewing angles $\alpha\sim 30-50$~deg. 
Around these angles, the
contribution from the O mode photons is dominant at high energies, and the
reduction of these photons through photon-axion conversion results in
a significantly cooler spectrum. As the viewing angle increases
further, the X-mode photons become increasingly dominant (see
Fig.~\ref{fig:fluxesB}), and the reduction of high energy photon
lessens. As a result, the effective temperature of the spectrum 
-- as measured through $E_{\rm peak}$ -- approaches the
value it would have without photon-axion conversion.  

To examine the dependence of our results on the effective
temperature $T_s$ of the star,
we consider the values $T_s=10^6$, $2.5\times 10^6$, and $5\times 10^6$~K
(where $T_s$ is the unredshifted temperature at the stellar surface)
for the magnetic field $B=4\times 10^{12}$~G.
We find that the shift in $E_{\rm peak}$ due to axion conversion
(compared to the case with equal temperature but no conversion) is
more pronounced at higher temperatures.  This is not
surprising given that the relative fraction of O to
X-mode photons is larger at higher temperatures.

The right panels of Fig.~\ref{fig:Aeff} show the phase-resolved
effective area
[i.e., $A_{\rm eff}(\alpha)\propto F(\alpha)/{E^4_{\rm peak}(\alpha)}$]
that an observer measures when fitting spectra
with a blackbody model.  To eliminate the
dependence of this diagnostic on the often poorly determined distance to the NS,
we normalize $A_{\rm eff} (\alpha)$ to its value near phase-on.
Since the photon-axion conversion probability is zero for precisely $\alpha=0$,
we normalize to a small angle $\alpha=2$~deg
which represents a ``phase-on'' spectrum as
an average of small angles around $\alpha=0$~deg.  When
photon-axion conversion is included, the variation of the effective
area with phase results from the interplay of two effects:
a decrease in $A_{\rm eff}$ due to flux dimming and
an increase in $A_{\rm eff}$ due to decreases in the peak energy and
inferred temperature.
Because of the $E_{\rm peak}^4$-dependence, the
latter effect dominates the former. This is particularly
the case for lower fields, which has the strongest
$E_{\rm peak}$ shift with phase.

\begin{figure*}[ht]
\includegraphics[scale=0.9,angle=0]{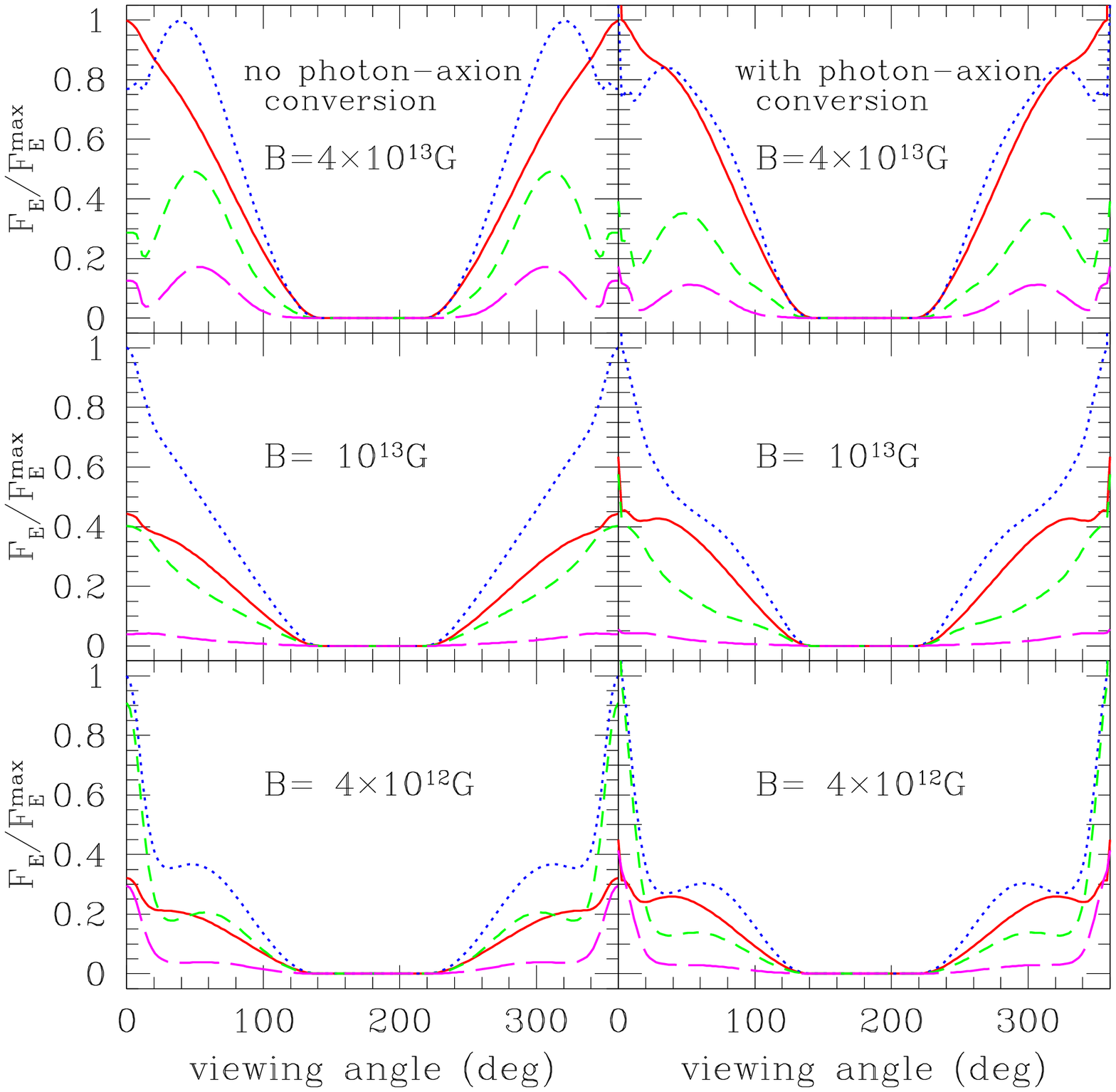}
\caption{Comparison of light curves at four energies with (right panels) and without (left panels) photon-axion
conversion.  The values of the energies are 0.5~keV {\em (solid curves)}, 1.5~keV {\em (dotted curves)},
2.5~keV {\em (dashed curves)}, 4~keV {\em (long-dashed curves)}. In each panel, light curves
are normalized to the maximum value among the energies considered.
\label{fig:lightcurves}}
\end{figure*}

The variation of inferred effective area with phase is a particularly
interesting and peculiar feature of photon-axion conversion. The
specific details of the variation, however, depend on NS and
axion parameters. In order to explore a wider region of 
$m_a-g_a$ parameter space, we compute the effective area for the
phase-averaged spectrum ($A_{\rm eff}^{\rm ave}$) with photon axion
conversion, normalized to that without axion conversion. 
This represents a measure of the ``correction'' that
photon-axion conversion makes to the inferred effective area (as
determined through measurements of phase-averaged flux and effective
temperature). For our purposes,
the phase-averaged spectrum of a hot spot provides an adequate approximation
for emission from the entire NS surface. The results of this
calculation are displayed in Fig.~\ref{fig:2D} for a combination of
magnetic fields and effective temperatures:
$B=4\times 10^{12}$~G, $T_s=5\times 10^6$~K; $B=4\times 10^{12}$~G, 
$T_s=2.5\times 10^6$~K; and $B=4\times 10^{13}$~G, 
$T_s=5\times 10^6$~K.  It is interesting to note that, in
all cases, there are regions of parameter space in which the inferred
effective areas are smaller/larger
than the uncorrected areas.
This is expected given the two counteracting effects described above.
The $m_a-g_a$ region shown in Fig.~\ref{fig:2D} is one for which,
at typical NS magnetic fields, there is a larger probability of
photon-axion conversion in the soft X-ray band (where NSs are routinely
observed).
The correction to the effective area becomes negligible for the largest
values of $m_a$ and ${g_a}^{-1}$ displayed in Fig.~\ref{fig:2D}.
This is expected since the photon-axion conversion probability tends to zero
for these values at the keV energies of interest here.
The results summarized by this figure have a strong potential
to constrain axion signatures for NSs with good distance
measurements (see the discussion in \S5.2).

In Fig.~\ref{fig:lightcurves}, we examine the effects of
photon-axion conversion on NS light curves at different
energies, for the same set of axion
parameters as in Fig.~\ref{fig:spectra}. The lightcurves are all normalized
to the maximum value of the flux among the set of energies considered
(0.5, 1.5, 2.5, and 4~keV). The main effect of photon-axion conversion
is the relative suppression of the flux at higher energies with respect
to that at lower energies. 

\begin{figure*}[ht]
\includegraphics[scale=0.9,angle=0]{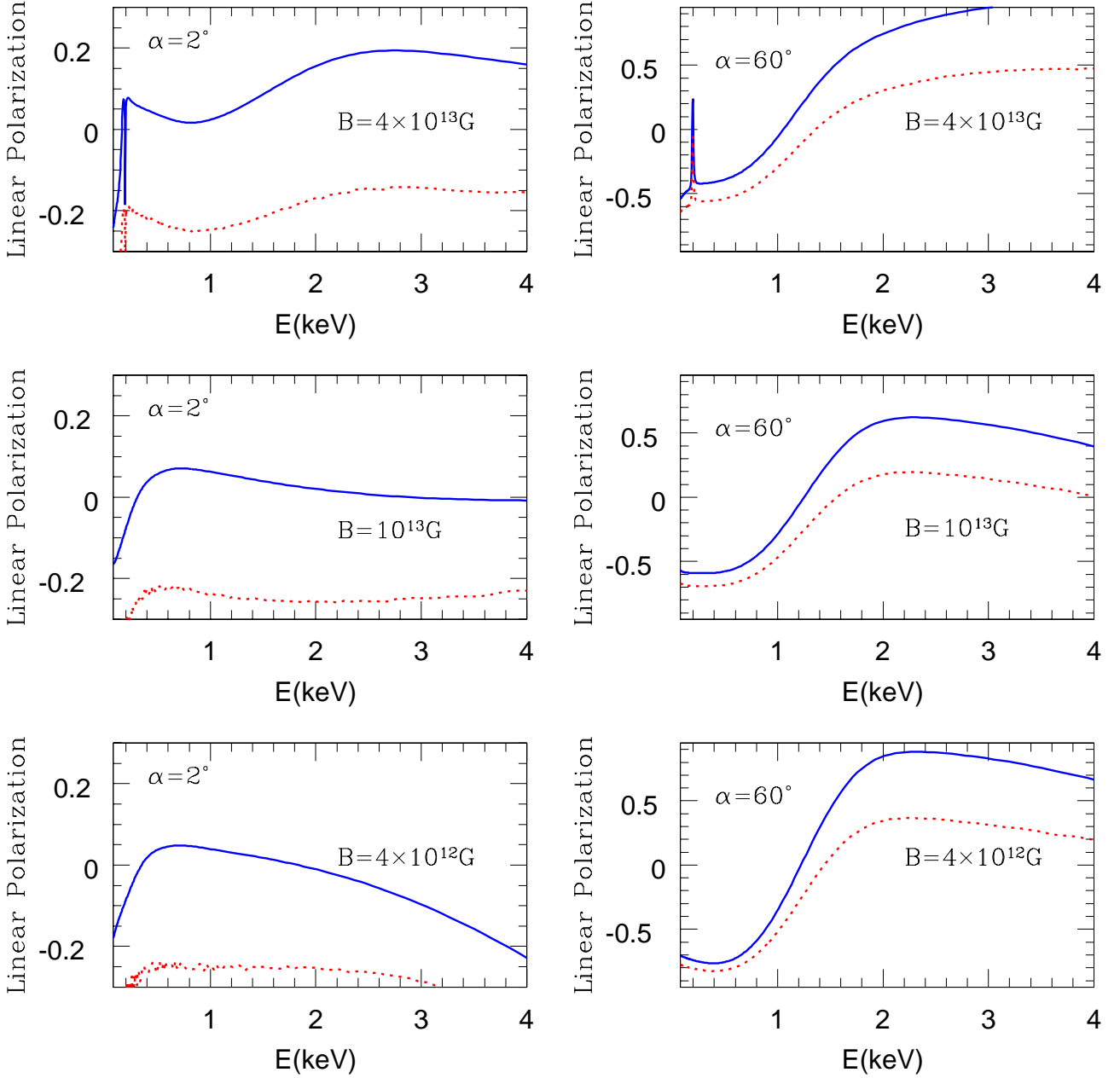}
\caption{Linear polarization, defined as $(F_O-F_X)/(F_O+F_X)$, 
for the cases with no photon-axion conversion {\em (solid lines)} and with
photon-axion conversion {(\em dotted lines)}. NS and axion parameters are the same
as in Fig.~2.
\label{fig:pol}}
\end{figure*}

Finally, we study the effect of photon-axion conversion on the
magnitude of linear polarization [see Eq.~(\ref{eq:pol})].  For
consistency with Fig.~\ref{fig:spectra}, we consider the same set of
parameters.  The polarization signal with and without conversion is
shown in Fig.~\ref{fig:pol}.  Since the main effect of axion
conversion is suppression of O-mode photons, the net result is a
reduction of the magnitude of linear polarization.  In addition, it is
noteworthy that the presence of photon-axion conversion causes the
plane of polarization to rotate from parallel (with respect to the plane
formed by the propagation direction and magnetic field)
to perpendicular for a phase-on spectrum (small $\alpha$, left panels
of Fig.~\ref{fig:pol}). This can be understood with reference to
Fig.~\ref{fig:spectra}: for a NS observed phase-on with respect to the
emission region, the intensity of the O mode photons slightly exceeds
the X-mode photon intensity. When photon-axion conversion is
considered, the O mode photon intensity is suppressed and becomes
smaller than the X-mode intensity, which results in an inversion of
the plane of polarization with respect to the case without
conversion. Clearly, the extent to which this occurs depends on the
specific axion parameter values, since the magnitude of the conversion
probability is a strong function of $m_a$ and $g_a$ (see
Fig.~\ref{fig:probs}).  For example, at $B=10^{13}$~G and $T_s=5\times
10^6$~K, the X-mode intensity of a phase-on spectrum is about 90\% of
the O-mode intensity at $E\sim 1$~keV. Therefore, if photon-axion
conversion occurs with probability $\ga 10\%$, it results in an
inversion of the plane of polarization. For the combination of axion
parameters adopted in Fig.~\ref{fig:probs}, a probability of $\ga
10\%$ is achieved when $g_a\ga 10^{-9}$~GeV$^{-1}$ (for $m_a=2\times
10^{-6}$~eV).  When $g_a=10^{-8}$~GeV$^{-1}$, the probability is $\ga
10\%$ for the entire range of axion masses considered ($7\times 10^{-7}-1.7\times
10^{-5}$~eV).

At larger viewing angles on the other hand (right panels of
Fig.~\ref{fig:pol}), it becomes evident one of the unique signatures
of vacuum polarization on the polarization signal of NSs with magnetic
fields $B\la 7\times 10^{13}$~G, that is the rotation at high energies
of the plane of polarization from perpendicular to parallel (Lai \& Ho
2003b).  Photon-axion conversion does not erase this feature; however,
as a result of the suppression of the O-mode photons which dominate at
higher energies, the shift of the plane occurs at higher energies than
it would without conversion. For the axion parameters adopted in
Fig.~\ref{fig:pol}, the energy at which the shift occurs with the
conversion differs by a factor of $\sim 15-40\%$ with respect to the
case with no conversion, with the precise value depending on the
magnetic field strength.

\subsection{Observational tests}
\label{subsection:observ}

Our theoretical results provide several methods for constraining axion
parameters from actual measurements of NS spectra.  The foremost
method is spectral analysis, often performed on phase-averaged spectra
when the resolution low. We showed in \S5.1 that the relative
suppression of the O-mode photons with respect to the X-mode photons
noticeably distorts the {\em spectral shape} with respect to the case
without conversion. This distortion is strongest in the high-energy
spectral tail, where the intensity of the O-mode photons is larger for
most viewing angles (see Fig.~\ref{fig:spectra}).  Observed
enhancements of high energy spectral tails are generally modeled with
a power law (often of unknown origin; e.g. Manzali et al. 2007, Durant
et al. 2011 for young and middle-aged pulsars; Perna et al. 2001,
Juett al. (2002), Rea et al. 2007 for magnetars).  However, a {\em
deficit} of high energy photons in a thermal fit could be a signature
of photon-axion conversion, and the axion parameters could then be
determined by means of detailed spectral fits that account for the
effect.  For this type of analysis, the distance to the NS is not
important since the main discriminant is the spectral shape, in
particular the deficit of high-energy photons.

For sources with good distance estimates\footnote{Distances to isolated
NSs are usually accurate to, at best, $\sim 10\%$, while the distances to NSs
in globular clusters are known to much higher precision.}
but too few X-ray counts to
perform detailed phase-resolved spectroscopy, Fig.~\ref{fig:2D}
demonstrates that the determination of the effective emission area (through 
measurements of flux and peak energy) can rule out large
regions of $m_a-g_a$ parameter space if the inferred NS radius is
close to the maximum or minimum value allowed by NS nuclear equations
of state (EOS).  The minimum radius constraint requires knowing that
the observed radiation is being emitted by the entire NS surface and
not from a small hot spot.  This may be the case when pulsations are
not detected or when the angle between the line of sight and magnetic
dipole axis is non-negligible and the measurement is made near pulse
minimum. The maximum constraint on the other hand does not require
other independent information. 

To illustrate how to use Fig.~\ref{fig:2D} with NS observations,
suppose $R_{max}$ is the maximum radius allowed by a particular EOS
(see, e.g., Lattimer \& Prakash 2007, for a review), and $A_{\rm max}$
is the corresponding surface area.  Then a NS whose inferred emission
area is $A\sim A_{\rm max}$ would rule out the $m_a-g_a$ region of the
parameter space for which the effective area 'correction' due to
photon-axion conversion is larger than 1. For $B=4\times 10^{12}$~G
and $T_s=5\times 10^6$~K (corresponding to a redshifted value $\sim
3.8\times 10^6$~K), the parameter space with $m_a\la 2\times
10^{-5}$~eV and ${g_a}^{-1}\la 4\times 10^{9}$~GeV (or ${g_a}\ga
2.5\times 10^{-8}$~GeV$^{-1}$) would be ruled out.  For NSs whose
radius determination relies on methods other than flux and peak energy
measurements\footnote{Examples of such methods of NS radius
  measurement (or more generally of the radius-to-mass ratio) include
  spectral lines (e.g. Cackett et al. 2009), thermal X-ray pulse
  profiles (e.g. Bogdanov et al. 2008; note that at low energies $E\la
  ~1$~keV, these pulse profiles are not very sensitive to the
  presence of photon axion conversion, as shown in Fig.~\ref{fig:lightcurves}),
  gravitational wave emission (Lenzi et al. 2009), quasi-periodic
  oscillations in accreting NSs (Miller et al. 1998), and neutrino
  emission from proto-stars (Lattimer \& Prakash 2007).  Masses can be
  independently inferred from radial velocity studies for NSs in
  binary systems (e.g. van Kerkwijk et al. 2011) and pulsar timing (a
  comprehensive review of all these methods can be found in Lattimer
  \& Prakash 2007).}, constraints on axion parameters can be obtained
using the method described above without any need for the radius to
be near the maximum or minimum allowed by the EOS.

As an aside, we note that the the effects of 
axions on the inferred area discussed above suggest 
that the use of NS effective radii to rule out nuclear EOSs may not be entirely reliable.
The existence of axions could severely bias such conclusions.

For sources with phase-resolved X-ray spectra, observations can be fit
with either blackbody or model atmosphere spectra
to obtain peak energies $E_{\rm peak}$
and effective areas $A_{\rm eff}$ as a function of rotation phase or
viewing angle $\alpha$.  Figure~\ref{fig:Aeff} shows a clear signature
of photon-axion conversion, i.e., a significant rise and fall of
effective area at intermediate viewing angles, especially for $B\lesssim
10^{13}\mbox{ G}$.  The important quantity for this diagnostic is the {\em relative} change
of the effective area with phase; thus, this method does not
require knowledge of the distance to the NS. It is also rather insensitive
to the NS radius/mass (i.e., surface gravity), since variations in this quantity produce
energy shifts that are the same at all phases, leaving 
the relative phase-dependent signatures of the conversion unaltered.
An exact estimate of the range of axion parameters that can be constrained
by means of this method cannot be made in absolute terms, since it clearly
depends on the quality of the data. However, from a theoretical
point of view, photon-axion signatures are imprinted in the NS spectra
as long as the conversion probability is non-negligible in the typical X-ray
band of observations. Fig.~\ref{fig:spectra} shows that this is the
case for an axion coupling strength $g_a\la 10^{-9}$~GeV$^{-1}$ and for
a large range of axion masses, up to $m_a\sim 10^{-4}$~eV if 
$g_a\la 10^{-8}$~GeV$^{-1}$. 

The theoretical framework and methods for extracting axion parameters
from observations of magnetic NSs is presented above.  Detailed
analyses of specific sources require a model
appropriate for the specific viewing and emission geometry, magnetic
field strength, and surface temperature for each source;
a comparison with the data can
then be performed by folding the model through the response function of the
detector. Such an analysis will be the subject of
future work.  Below, we briefly review NSs that serve as potential
candidates for obtaining constraints on axion physics.  These sources
possess bright X-ray thermal emission from the NS
surface.  Among the observed NS population, such sources can be
grouped into several classes (which may not be mutually exclusive
groups; see, e.g., Kaspi~2010; Perna \& Pons~2011; Pons \&
Perna~2011).  One class is the magnetars, which generally possess
superstrong magnetic fields
$B\ga 10^{14}\mbox{ G}$.\footnote{Note however that the bursting
  behaviour used to characterize magnetars has also been seen
  in a lower magnetic field object (see Rea et al 2010).}
For these objects, vacuum resonance effects lead to 
emission that is predominantly in the X-mode.  Therefore the possible
conversion of O-mode photons to axions will not produce an observable
effect on the total radiation spectrum.  Another class of NSs are the
rotation-powered pulsars; these are the classic radio pulsars that are
powered by the loss of rotational energy due to emission of
electromagnetic radiation.  These NSs can have magnetic fields up to
the magnetar regime $B\la 10^{14}\mbox{ G}$, and a small fraction of
sources are sufficiently bright in the X-rays (see, e.g., Pavlov et
al.~2002; Mereghetti~2007; Zavlin~2009, for reviews).  These NSs
are good candidates if their X-ray emission is not
dominated by magnetospheric emission (as in, e.g., PSR~B0656$+$14).

The isolated NSs are another interesting class.  The seven confirmed
isolated NSs have $B\approx (1-4)\times 10^{13}\mbox{ G}$ inferred
from timing measurements (Kaplan \& van Kerkwijk~2009, 2011), while
features in their X-ray spectrum suggest $B\approx (5-10)\times
10^{13}\mbox{ G}$ if the features are due to the proton cyclotron
resonance (Haberl~2007).  We note that the observed X-ray spectra of
isolated NSs are generally well-fit (except around spectral lines) by
blackbodies, including Wien-like behavior at high energies (van
Kerkwijk \& Kaplan 2007).  Various explanations have been proposed to
explain the soft high energy tails as compared to the hard tails
predicted by atmosphere models (e.g., Romani 1987), such as emission
from a partially optically thin atmosphere (e.g., Motch et al. 2003)
or the effect of vacuum polarization (see \S2 and Ho \& Lai 2003).
Here we have shown that photon-axion conversion can suppress
high-energy emission and produce softer tails (see
Fig.~\ref{fig:spectra}).  However, this cannot explain the spectra of
isolated NSs.  These NSs have surface temperatures $\la 10^6$~K, and,
as we discussed in Sec.~2, the conversion is less effective at low
temperatures due to the low fraction of O-mode photons. However, if
soft tails (softer than the blackbody) were detected in NSs with
relatively high surface temperature, then photon-axion conversion
could indeed be a competing explanation.

A final class is the central compact object neutron
stars (CCOs), which have $B\sim 10^{10}-10^{11}\mbox{ G}$, obtained
from timing and spectral analyses (see, e.g., Halpern \&
Gotthelf~2010).  At these low fields, the vacuum resonance is outside
of the atmosphere, and the emerging intensity has a high fraction of
O-mode photons; their surface temperatures are
constrained to be $\la (2-3)\times 10^6$~K, while hot spot
temperatures are quite high [$\sim (4-6)\times 10^6$~K].  These
characteristics, combined with the fact that the photon-axion
conversion probability at these lower fields is still high for a wide
range of axion parameters (Jimenez et al. 2011), make CCOs
potentially good candidates.  Current data on many of the NSs
belonging to the classes described above are already of
sufficient quality to allow axion constraints to be derived.

Finally, we note that X-ray polarization measurements, like those that
will be performed by the forthcoming {\it Gravity and Extreme
  Magnetism Small Explorer} (GEMS) satellite (Jahoda 2010),
will increase the robustness of axion constraints, as well
as provide an independent measure of the stellar magnetic field
strength and geometry (van Adelsberg \& Perna 2009).

\section{Summary}

Constraints on the axion mass and coupling strength by
means of observations of magnetized objects have been discussed in
the recent literature (e.g., Lai \& Heyl 2006; Chelouche et al. 2009;
Gill \& Heyl 2011; Jimenez et al. 2011) as a means to
complement and independently test constraints obtained from other
methods. The studies above emphasized different
observational aspects and tests.  The modification to the X-ray
spectra of highly magnetized NSs induced by photon-axion conversion
was discussed by Lai \& Heyl (2006) using a blackbody spectrum to
model the photon spectrum from NSs.
Chelouche et al. (2009) discussed spectral features produced
in the sub-mm/IR wavelength regime for magnetized NSs,
while Gill \& Heyl (2011) considered limits that can be derived
from polarization measurements of white dwarfs.  Jimenez et al. (2011)
focused their analysis on constraints that can be obtained through
observations of eclipsing white dwarfs in binaries, but they also
discussed qualitatively the potential of NSs.  Since most of the
photon-axion conversion happens at a distance of many stellar radii from the NS, 
they pointed out that there are two configurations that should
be considered: light from a background object passing through the
magnetic field of the NS (i.e., occultation) and a binary containing a
NS where the companion transits close enough to the line
of sight for its light to be influenced by the NS magnetic field.  In
these two cases the treatment is much simpler as the physics of
photon propagation in the NS atmosphere is irrelevant.  However, they
determined that the occultation probability is too low to be
astrophysically relevant and that there are no known binary
systems involving a NS that are detached enough to yield a clean
constraint.

Since NSs are routinely observed in the X-ray domain, the goal of our
paper has been an extensive exploration of the actual constraints on
axion physics that can be obtained from observations of NS thermal
spectra.  We have also explored what can be learned from future X-ray
polarization measurements. For these purposes, we used detailed,
magnetized atmosphere models, which properly account for the energy
and angle-dependent emerging intensities of the two polarization
modes, including mode conversion due to vacuum polarization
effects. The emergent O mode intensity was coupled with the energy and
angle-dependent photon-axion conversion probability, and this allowed
us to make theoretical predictions for phase-resolved spectra with
photon axion-conversion in the relevant regime of magnetic fields and
temperatures for NSs.  Since the relative intensities of the O and X
mode photons depend strongly on the angle at which they emerge with
respect to the magnetic field direction, we considered emission from a
region consisting of a hot spot with an axis coincident with the NS
magnetic field axis.  { Thermal emission from most X-ray bright NSs
  does indeed indicate that we are observing hot spots (this from both
  non-negligible pulse fractions and small inferred emission areas).
  In order to extend axion studies by means of comparison with
  observations of {\em any} NS (especially those emitting from the
  entire surface, with non-dipolar, complicated B topologies, and fast
  rotators), one would have to include, locally, magnetic fields
  non-normal to the surface in the atmospheres, investigate how
  non-dipolar fields influence both the emission spectra as well as
  the photon-axion conversion, and, for the polarization, explicitly
  integrate the Stokes parameters to $R_{pl}$.  However, since the
  magnetic field structure over the entire surface of a NS is not
  fully known apriori, determination of the axion parameters would be
  more degenerate under these more general conditions. }

Our analysis has identified some features in NS spectra
that bear the telltale signs of photon-axion conversion: 

\begin{itemize}

\item The NS {\em spectral shape} is noticeably distorted compared
  to the case without photon-axion conversion, with suppression of the
  high-energy tail for viewing angles 
$\sim 20-70$~deg (with respect to the center of the emission region).  
Detailed spectral fits can yield axion parameters.
  This analysis does not require knowledge of the distance to the NS,
  as the axion signatures are imprinted in the
  spectral shape.

\item 
The spectral suppression of the O-mode photons by photon-axion conversion
dominates the high energy tail of the spectrum for a range of
viewing angles.  In addition, it results in a shift of the peak energy, $E_{\rm peak}$,
towards lower values for a range of rotation phases (i.e., 
viewing angles).  As a result, the effective area of the spot $A_{\rm
  eff}$ (as inferred from measurements of the flux and peak
spectral energy) is larger than its value without
axion conversion. We demonstrated that a clear
signature of photon-axion conversion is a significant rise and
fall of effective area (normalized to the phase-on value) at
intermediate viewing angles, especially for $B\lesssim 10^{13}\mbox{G}$.
Such a measurement requires high signal-to-noise phase-resolved
spectra, but does not require knowledge of the distance to the
NS, since the axion signature appears in the {\em relative}
change of the effective area with phase.
 
\item For a star emitting from the entire surface, photon-axion
  conversion can substantially affect the inferred NS emission area
  by $\lesssim 50\%$, and the inferred radius by $\lesssim 20\%$.
  These values are measured using the flux and peak
  energy of the thermal spectrum. If the distance is well
  constrained, and the NS radius is known through different methods,
  then the inferred emission area can be directly translated into a
  constraint in the $m_a-g_a$ plane. If the NS
  radius is not independently known, then an inferred emission area that
  exceeds the maximum value allowed by NS equations of state could indicate
  the presence of photon-axion conversion.

\item In the absence of photon-axion conversion, radiation from a hot spot 
  observed phase-on at energies $\sim
  2-3$~keV is linearly polarized in the plane formed by ${\bf B}$ and
  the direction of photon propagation.  Conversion 
  rotates the plane of polarization and leads to radiation
  polarized perpendicular to the plane of ${\bf B}$ and the
  direction of photon propagation, for a range of axion parameters.
\end{itemize}

We concluded \S5.2 with a discussion of appropriate sources to study within the 
theoretical framework developed here for probing axions.  While this paper has 
outlined the main
elements for connecting theoretical ideas of photon-axion conversion
to {\em actual} observations that are made of NSs, studying each
suitable source will require a specific suite of models, tailored
to the particular object (e.g., its magnetic field, surface
temperature, emission and viewing geometry). The theoretical models are then convolved with the
detector response, and a process of minimization identifies the most likely values of the 
parameters.
Detailed analysis of the most promising NS candidates will be the subject of
future work.

\acknowledgements This work was partially supported by grants NSF
AST-1009396, AR1-12003X, and DD1-12053X (RP); WCGH is supported by
STFC in the UK; LV and RJ are supported by MICINN grab AYA 2008-03531.
RP, RJ and LV thank the Centro de Ciencias de Benasque Pedro Pascual
where this work was initiated. We further acknowledge Carlos
Pe\~na-Garay, Dong Lai, Jeremy Heyl and Enrique Fernadez-Martinez for
insightful discussions, and an anonymous referee for helpful
comments on our manuscript.  WCGH appreciates the use of the computer
facilities at the Kavli Institute for Particle Astrophysics and
Cosmology.


\begin{thebibliography}{}

\bibitem[Adler (1971)]{adler71} Adler, S. L. 1971, Ann. Phys., 67, 599
\bibitem[Alcock \& Illarionov (1980)]{alcock80} Alcock, C. \& Illarionov, A. 1980, ApJ, 235, 534
\bibitem[Andriamonje et al. (2007)]{Andria} Andriamonje, S. et al. 2007, J. Cosm. Astropart. Phys., 4, 10
\bibitem[Arvanitaki et al. (2009)]{arva09}Arvanitaki, A.,
Dimopoulos, S., Dubovsky, S., Kaloper,~N.,  March-Russell, J. 2010, Phys. Rev., D, 81, 123530.
\bibitem[Asztalos et al. (2004)]{asz04}Asztalos, S. J.  et al., 2004, Phys. Rev. D, 69, 011101   
\bibitem[Asztalos et al. (2010)]{asz10}Asztalos, S. J.  et al., 2010, Phys. Rev. Lett. 104, 041301 
\bibitem[Avgoustidis et al. (2010)]{Tasos} Avgoustidis,~A. Burrage,~C.  Redondo,~J., Verde,~L., Jimenez,~R. 2010,
  JCAP, 1010, 24, preprint arXiv:1004.2053
\bibitem[Bahcall (1989)]{bah89} Bahcall, J. 1989, Neutrino Astrophysics, (Cambridge: Cambridge Univ. Press)
\bibitem[Bernardini et al. (2011)]{Ber11} Bernardini, F., Perna, R., Gotthelf, E., Israel, G. Rea, N., Stella, L.
2011, MNRAS, 418, 638
\bibitem[Bogdanov et al. (2008)]{bog08} Bogdanov, S., Grindlay, J. E., Rybicki, G. B. 2008, ApJ, 689, 407	
\bibitem[Brown et al. (2002)]{brown02} Brown, E. F., Bildsten, L., \& Chang, P. 2002, ApJ, 574, 920
\bibitem[Cackett et al. (2008)]{cack08}Cackett, E. M., Miller, J. M., Bhattacharyya, S., Grindlay,
J. E., Homan, J., van der Klis, M., Miller, M. C., Strohmayer, T. E., Wijnands, R. 2008, ApJ, 674, 415
\bibitem[Chang et al. (2010)]{chang10} Chang, P., Bildsten, L., \& Arras, P. 2010, ApJ, 723, 719
\bibitem[Chelouche et al. (2009)]{che09} Chelouche, D., Rabadan, R., Pavlov, S. S., Castejon, F. 2009, ApJ, 180, 1
\bibitem[De Panlis et al. (1987)]{Depanlis} De Panlis, S.  et al. 1987, Phys. Rev. Lett. 59, 839  
\bibitem[Durant et al. (2011)]{dur11} Durant, M., Kargaltsev, O., Pavlov, G. G., 2011, arXiv e-print (arXiv:1109.1984)
\bibitem[Eidelman et al. (2004)]{SN} Eidelman, S.  {\it et al.} 2004, Phys.\ Lett.\,  B592, 1 
\bibitem[Gill \& Heyl(2011)]{Heyl} Gill, R., \& Heyl, J.~S.\ 2011, arXiv:1105.2083
\bibitem[Haberl (2007)]{Hab} Haberl, F. 2007, Astrophys. and Space Science, 308, 181
\bibitem[Halpern \& Gotthelf (2010)]{HG10} Halpern, J. P., Gotthelf, E. V. 2010, ApJ, 709, 436
\bibitem[Haxton (1995)]{hax05} Haxton, W. C. 1995, ARA\&A, 33, 459
\bibitem[Heyl \& Shaviv (2000)]{hs00} Heyl, J. S., Shaviv, N. J. 2000, MNRAS, 311, 555
\bibitem[Heyl \& Shaviv (2002)]{hs02} Heyl, J. S., Shaviv, N. J. 2002, Phys. Rev.D, 66, 023002
\bibitem[Heyl et al. (2003)]{hsl03} Heyl, J. S., Shaviv, N. J., Lloyd, D. 2003, MNRAS, 342, 134
\bibitem[Ho \& Lai (2001)]{Ho01} Ho, W. C. G. \& Lai, D. 2001, MNRAS, 327, 1081 
\bibitem[Ho \& Lai (2003)]{Ho03a} Ho, W. C. G. \& Lai, D. 2003, MNRAS, 588, 962 
\bibitem[Ho et al. (2008)]{Ho08} Ho, W. C. G., Potekhin, A. Y., Chabrier, G. 2008, ApJS, 178, 102
\bibitem[Jahoda (2010)]{jah10} Jahoda, K. 2010, Proc. SPIE, 7732, 77320W
\bibitem[Juett et al. (2002)]{ju02} Juett, A. M., Marshall, H. L., Chakrabarty, D., Schulz, N. S
2002, ApJ, 568L, 31
\bibitem[Jimenez et al.(2011)]{RJ} Jimenez, R., Pe{\~n}a-Garay, C., \& Verde, L.\ 2011, Physics Letters B, 703, 232
\bibitem[Kaplan \& van Kerkwijk (2009)]{KK09}   Kaplan, D. L., van Kerkwijk, M. H. 2009, ApJ, 705, 798
\bibitem[Kaplan \& van Kerkwijk (2011)]{KK11}   Kaplan, D. L., van Kerkwijk, M. H. 2011, ApJ, 740, 30
\bibitem[Kaspi (2010)]{Kaspi} Kaspi, V. M. 2010, Publ. of the Nat. Acad. of Science, 107, 7147
\bibitem[Lai \& Heyl (2006)]{LH06} Lai, D. \& Heyl, J. 2006, Phys. Rev. D, 2006, 74, 13003
\bibitem[Lai \& Ho (2002)]{Lai02} Lai, D. \& Ho, W. C. G. 2002, ApJ, 566, 373
\bibitem[Lai \& Ho (2003a)]{Lai03} Lai, D. \& Ho, W. C. G. 2003a, ApJ, 588, 962
\bibitem[Lai \& Ho (2003b)]{Lai03b} Lai, D. \& Ho, W. C. G. 2003b, Phys. Rev. Lett., 91, 071101
\bibitem[Lattimer \& Prakash (2007)]{Lat07} Lattimer, J. M., \& Prakash, M. 2007, Phys. Rep., 442, 1
\bibitem[Lenzi et al. (2009)]{len09} Lenzi, C. H., Malheiro, M., Marinho, R. M., Marranghello, G. F.,
Providencia, C.  2009, Journ. of Phys. 154, 1
\bibitem[Manzali et al. (2007)]{man07}  Manzali, A., De Luca, A., Caraveo, P. A. 2007, ApJ, 669, 570
\bibitem[Mereghetti (2007)]{Me07} Mereghetti, S. 2007, MmSAI, 78, 644
\bibitem[Miller et al. (1998)]{mil98} Miller, M. C., Lamb, F.K., Psaltis, D. 1998, ApJ, 508, 791 
\bibitem[Mori \& Ho (2007)]{mori07} Mori, K. \& Ho, W. C. G. 2007, MNRAS, 377, 905
\bibitem[Motch et al. (2003)]{mot03} Motch, C., Zavlin, V. E., Haberl, F. 2003, A\&A, 408, 323
\bibitem[Page (1995)]{pag95} Page, D. 1995, ApJ, 442, 273
\bibitem[Pavlov et al. (2002)]{pav02} Pavlov, G. G., Sanwal, D., Garmire, G. P., Zavlin, V. E.  2002, Neutron Stars in Supernova
  Remnants, ASP Conference Series, Vol. 271, held in Boston, MA, USA,
  14-17 August 2001. Ed. Patrick O. Slane and Bryan
  M. Gaensler. San Francisco: ASP, p.247
\bibitem[Pavlov et al. (1994)]{pav94} Pavlov, G. G., Shibanov, Yu. A., Ventura, J., Zavlin, V. E. 1994, A\&A, 289, 837
\bibitem[Pavlov \& Zavlin (2000)]{pav00} Pavlov, G. G. \& Zavlin, V. E. 2000, ApJ, 529, 1011
\bibitem[Peccei \& Quinn (1977)]{PQ} Peccei, R.~D., Quinn, H.~R. 1977,  Phys.\ Rev.\ Lett., 38, 1440
\bibitem[Pechenik et al. (1983)]{Pec83} Pechenick, K. R., Ftaclas, C., Cohen, J. M. 1983, ApJ, 274, 846
\bibitem[Perna et al. (2001)]{pe01} Perna, R., Heyl, J. S., Hernquist, L. E., Juett, A. M., Chakrabarty, D.
2001, ApJ, 557, 18
\bibitem[Perna \& Gotthelf (2008)]{PG08} Perna, R. \& Gotthelf, E. V. 2008, ApJ, 681, 522 
\bibitem[Perna \& Pons (2011)]{PP11a} Perna, R. \& Pons, J. A. 2011, ApJ, 727, L51
\bibitem[Pons \& Perna (2011)]{PP11b} Pons, J. A. \& Perna, R. 2011, ApJ, 741, 123
\bibitem[Potekhin et al. (2004)]{Pot04} Potekhin, A. Y., Lai, D., Chabrier, G., Ho, Wynn C. G. 2004, ApJ, 612, 1034
\bibitem[Preskill et al. (1983)]{Cosmo} Preskill, J. Wise, M. B., Wilczek, F. 1983,  Phys.\ Lett.\,  B, 120, 127
\bibitem[Raffelt (2008)]{Raff08} Raffelt, G. G. 2008, Lect. Notes Phys., 741, 51
\bibitem[Raffelt \& Stodolski (1988)]{RS88} Raffelt, G. \& Stodolski, L. 1988, Phys Rev. D, 37, 1237
\bibitem[Rea et al. (2007)]{rea07} Rea, N. et al. 2007, MNRAS, 381, 293
\bibitem[Rea et al. (2010)]{Rea10} Rea, N. et al. 2010, Science, 330, 944
\bibitem[Romani (1987)]{Rom87} Romani, R. W. 1987, ApJ, 313, 718
\bibitem[Suleimanov et al. (2009)]{sul09} Suleimanov, V., Potekhin, A. Y., \& Werner, K. 2009, A\&A, 500, 891
\bibitem[Tsai \& Erber (1975)]{tsai75} Tsai, W. Y. \& Erber, T. 1975, Phys. Rev. D, 12, 1132
\bibitem[van Adelsberg \& Lai (2006)]{van06} van Adelsberg, M. \& Lai, D. 2006, MNRAS, 373, 1495
\bibitem[van Adelsberg \& Perna (2009)]{van09} van Adelsberg, M. \& Perna, R. 2009, MNRAS, 399, 1523
\bibitem[van Kerkwijk \& Kaplan (2007)]{vank07} van Kerkwijk, M. H. \& Kaplan, D. L. 2007, Ap\&SS,
 308, 191
\bibitem[van Kerkwijk et al. (2011)]{vank11} van Kerkwijk, M. H., Breton, R. P., Kulkarni, S. R. 2011, ApJ, 728, 95
\bibitem[Wuensch et al. (1989)]{wuen} Wuensch, W. U.  et al., 1989, Phys. Rev., D, 40, 3153
\bibitem[Zavlin (2009)]{zav} Zavlin, V. E. 2009, in Neutron Stars and Pulsars,
  Astrophysics and Space Science Library, Vol. 357. Springer Berlin Heidelberg, p. 181

\end{thebibliography}
\end{document}